\documentclass[12pt]{article}
\usepackage{url}
\input{epsf}

\font\bm=cmmib10 at 10pt
\font\bms=cmmib10 at 7pt \textfont9=\bm \scriptfont9=\bms
\usepackage{amsfonts}
\usepackage{multirow}
\mathchardef\balpha= "790B
\mathchardef\bbeta= "790C
\mathchardef\bTheta= "7902
\mathchardef\bzeta= "7910
\mathchardef\bOmega= "790A
\mathchardef\bGamma= "7900
\mathchardef\bDelta= "7901
\mathchardef\bPhi= "7908
\mathchardef\bphi= "791E
\mathchardef\bomega= "7921
\mathchardef\bxi= "7918
\mathchardef\bet= "7911
\mathchardef\brho= "791A
\mathchardef\btau= "791C
\mathchardef\bmu= "7916
\mathchardef\bvarpi= "7924

\def \lvec{(\kern-.26em(}
\def\pmb#1{\setbox0=\hbox{#1}%
\def \lvec{(\kern-.26em(}
\kern-.025em\copy0\kern-\wd0
\kern.05em\copy0\kern-\wd0
\kern-.025em\raise.0433em\box0 }
\mathchardef\btheta= "7912
\usepackage{amsmath}
\usepackage{authblk}

\usepackage{graphicx}
\usepackage{dcolumn}

\begin{document}

\title{Impact of Changing Greenhouse Gas Concentrations on Ontario's Climate}
\author{W. A. van Wijngaarden}
\affil{Department of Physics and Astronomy, York University, Canada, wavw@yorku.ca}
\renewcommand\Affilfont{\itshape\small}
\date{\today}
\maketitle

\noindent The effect of changing greenhouse gas concentrations, most notably carbon dioxide, CO$_2$, on climate was examined.  In particular, calculations of the climate sensitivity, the warming of the Earth due to a doubling of atmospheric carbon dioxide, are discussed.  Greenhouse gas concentrations, as determined from air bubbles trapped in ice as well as at Mauna Loa, Hawaii are presented.  The greenhouse gas amounts generated by Canada and Ontario were used to estimate their respective contributions to global warming.  Ontario was responsible for only 0.35\% of the world's CO$_2$ emissions in 2019 and this amount was 20\% lower than in 2005.  The predictions of Global Climate Models (GCMs) regarding temperature, polar ice caps, oceans, precipitation and extreme events were compared to observations.  Records since 1880 show an overall warming of about 1 $^o$C.  However, the GCMs do not account for observed decadal temperature fluctuations and consistently overestimate the warming.  Ontario's contribution to global warming is only $9.2 \times 10^{-5}$ $^o$C/year using the Intergovernmental Panel on Climate Change (IPCC) recommended climate sensitivity value.  Measurements of the polar ice caps reveal a decrease in the minimum September Arctic sea ice extent during 1979-2022 but the trend levelled off after 2007; while the average Antarctic sea ice extent slightly increased.  Sea level increased slightly throughout the 20th century.  Ontario's contribution to anthropogenic sea level rise is about 0.005 mm/year.  Sea level along Ontario's Hudson Bay coast is decreasing due to isostatic rebound of the land following the last Ice Age.  The change to ocean acidity due to CO$_2$ absorption from the atmosphere is negligible compared to that due to tides, ocean depth and seasonal effects.  Ontario's contribution to ocean acidification is estimated to be $6 \times 10^{-6}$ pH/year.  No changes in precipitation in North America over the 19th and 20th centuries, nor at Toronto since 1843, were found.  The Great Lake levels are remarkably constant over the past century showing no evidence of a change in the incidence of flooding.  No evidence was found that the frequency of extreme events such as hurricanes or tornadoes increased during recent decades.  The number of forest fires in Canada and Ontario decreased during 1990 to 2020.

%

\newpage
\section{Climate}
\subsection{Introduction}
In March 2019, the United Nations General Assembly President Maria Garces warned that only 11 years remain to avert catastrophic climate change stating ``we are the last generation that can prevent irreparable damage to our planet'' \cite{Garces}.  These concerns were amplified by the latest report issued by the Intergovernmental  Panel on Climate Change (IPCC) in August, 2021.  ``It has been clear for decades that the Earth's climate is changing, and the role of human  influence on the climate system is undisputed'' said V. Masson-Delmotte, Co-Chair of the IPCC Working Group \cite{IPCC1990,IPCC1996,IPCC2001,IPCC2007,IPCC2014,IPCC2021}.  
The IPCC report warns that increasing greenhouse gases due to the combustion of fossil fuels have not only caused global warming, but changed the Earth's ice cover, precipitation, sea level and ocean acidity as well as affected climate extreme events such as cyclones \cite{IPCC2021}.    

\subsection{Historical Climate}
\begin{figure}\centering
	\includegraphics[height=60mm,width=.5\columnwidth]{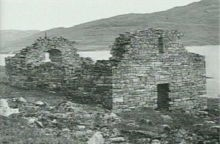}
	\caption{Remains of Hvalsey Church located in the Eastern Viking Settlement of Greenland \cite{Hvalsey}.  The Vikings settled Greenland during the Medieval Warm Period which extended from about 900 to 1300 A.D.   Cooling temperatures in the late 1300s are believed to have led to the demise of these settlements.  This illustrates the significance of natural climate change.
		\label{VikingKerk}}
\end{figure}

The first question to ask when considering climate change is whether the Earth's climate is constant.  The occurrence of Ice Ages indicates the answer is an emphatic No.  Ice ages are caused by small changes in the Earth's orbit around the sun known as Milankovitch cyles that have periods of about 100,000 years \cite{Milan}.  The next question is whether the climate has remained constant throughout recorded history.  Temperatures in the North Atlantic were warmer during the so called Medieval Warm Period from 900 to about 1300 A.D.  This allowed the Vikings to settle in southern Greenland and farm with sheep and goats!  Two settlements were established which at their peak had a total population of several thousand inhabitants \cite{Brown2000, Lynnerup2000}.  Trade with Europe was brisk in walrus ivory for several hundred years until the early 14th century.  It appears there was a shift to a markedly cooler climate in the Northern Hemisphere in the 1300s that led to the demise of the Viking settlements.  Radiocarbon dating of plant material found beneath glaciers on Baffin Island and Iceland shows glaciers began to grow in the late 1300s \cite{Miller2012}.  The stone ruins of a Church at Hvalsey, Greenland shown in Fig. \ref{VikingKerk} are a poignant reminder of the power of natural climate change.

The period from 1400 to 1700 is known as the Little Ice Age.  Temperatures in Europe were cooler.  Swiss villagers even offered prayers to stop advancing glaciers from destroying their alpine villages \cite{Cowie2007,Squires2009}.  Rivers such as the Thames and Rhine regularly froze during winter, something that has seldom occurred in the last 150 years.  The Little Ice Age ended around 1700.  The reasons for the Medieval Warm Period and Little Ice Age are not well understood.  An obvious possibility is variation of the amount of heat produced by the sun.  Satellite measurements of the solar intensity only exist from 1975 onwards \cite{ACRIM2012}.  These NASA data show the sun's intensity varies by about 0.1\% in a cyclic fashion having a period of 11 years.  This coincides with the well known sunspot cycle recorded over the last 200 years.   Sunspots were first observed in about 1600.  The sharp decreases in sunspot number in the 1600s and immediately after 1800 are known as the Maunder and Dalton Minima, respectively \cite{Sunspot}.  Possibly, the sun had fewer sunspots during the 1600s (as opposed to a lack of scientists studying sunspot activity) which corresponded to a reduced solar intensity that contributed to the Little Ice Age.

\subsection{Greenhouse Effect}
\begin{figure}\centering
	\includegraphics[height=80mm,width=.8\columnwidth]{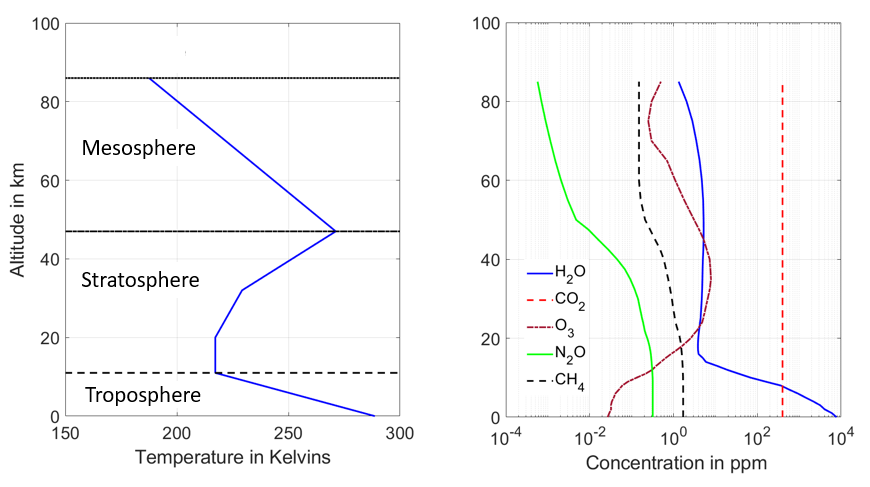}
	\caption{{\bf Left.} A midlatitude atmospheric temperature profile \cite{Temp}. The Earth's mean surface temperature is 15.5 degrees Celsius or 288.7 Kelvins.  {\bf Right.} Observed concentrations of greenhouse molecules in units of parts per million versus altitude \cite{Anderson}. 
		\label{GGNT}}
\end{figure}

The atmosphere is essential for making Earth habitable.  Dry air consists of 78\% Nitrogen, 21\% Oxygen molecules and 1\% Argon atoms.  Carbon dioxide (CO$_2$), Methane (CH$_4$) and Nitrous Oxide (N$_2$O) are only found in trace amounts with average surface concentrations in 2021 of 415, 1.9 and 0.32 parts per million (ppm) respectively.  Water vapour is a very important atmospheric constituent.  Its concentration varies considerably from as high as 4\% in the warm tropics at high humidity to minuscule amounts in the polar regions during winter.  Fig. \ref{GGNT} shows the observed concentration dependence of these trace gases on altitude \cite{Anderson}.

\begin{figure} \centering
	\includegraphics[height=60mm,width=.8\columnwidth]{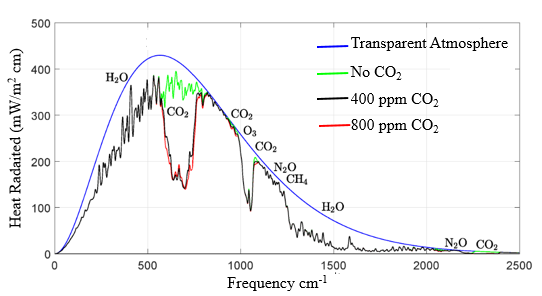} 
	\caption{Effects of carbon dioxide concentration on heat radiated by Earth and its atmosphere for a clear sky \cite{vW2020}. The smooth blue line is heat that would be radiated from a surface at temperature 15.5 Celsius or 288.7 Kelvins for a transparent atmosphere having no greenhouse gases.  The three other lines consider the case of H$_2$O, O$_3$, CH$_4$ and N$_2$O as well as: no CO$_2$ (green line), 400 ppm CO$_2$ (black line) and 800 ppm CO$_2$ (red line).  The difference between the red and black curves is what all the global warming fuss is about.
		\label{CO2}}
\end{figure}

The temperature of the Earth's atmosphere for a midlatitude position, where Ontario is located, is shown in Fig. \ref{GGNT} \cite{Temp}.  Weather events occur in the lowest part of the atmosphere called the troposphere.  The temperature is observed to decrease rapidly from an average surface temperature of about 15.5 degrees Celsius or 288.7 Kelvins until the tropopause which at midlatitudes occurs at an altitude of 11 km.  The temperature increases in the stratosphere located from 11 to 47 km, due to absorption of ultraviolet sunlight by ozone molecules.  The temperature drops sharply in the mesosphere located above 47 km.  

All objects having a finite temperature radiate energy.  The sun is very hot and therefore produces visible light.  Cooler objects such as the Earth's surface and atmosphere generate heat or infrared radiation.  
This infrared radiation exists at a wide range of infrared colours, which scientists refer to as frequencies or wavelengths, that the human eye cannot see.  In the absence of clouds, most sunlight is transmitted through the atmosphere.  In contrast, infrared radiation, is strongly absorbed by the atmosphere.   A gas that does not absorb visible sunlight but strongly absorbs infrared radiation is called a greenhouse gas.  

In the absence of greenhouse gases, radiation emitted by the Earth's surface would pass through a transparent atmosphere to outer space.  Greenhouse gases absorb the radiation which is then reemitted in all directions, some upwards and some back to the surface, warming it.  This is called the Greenhouse Effect.  

\begin{figure} \centering
\includegraphics[height=80mm,width=1\columnwidth]{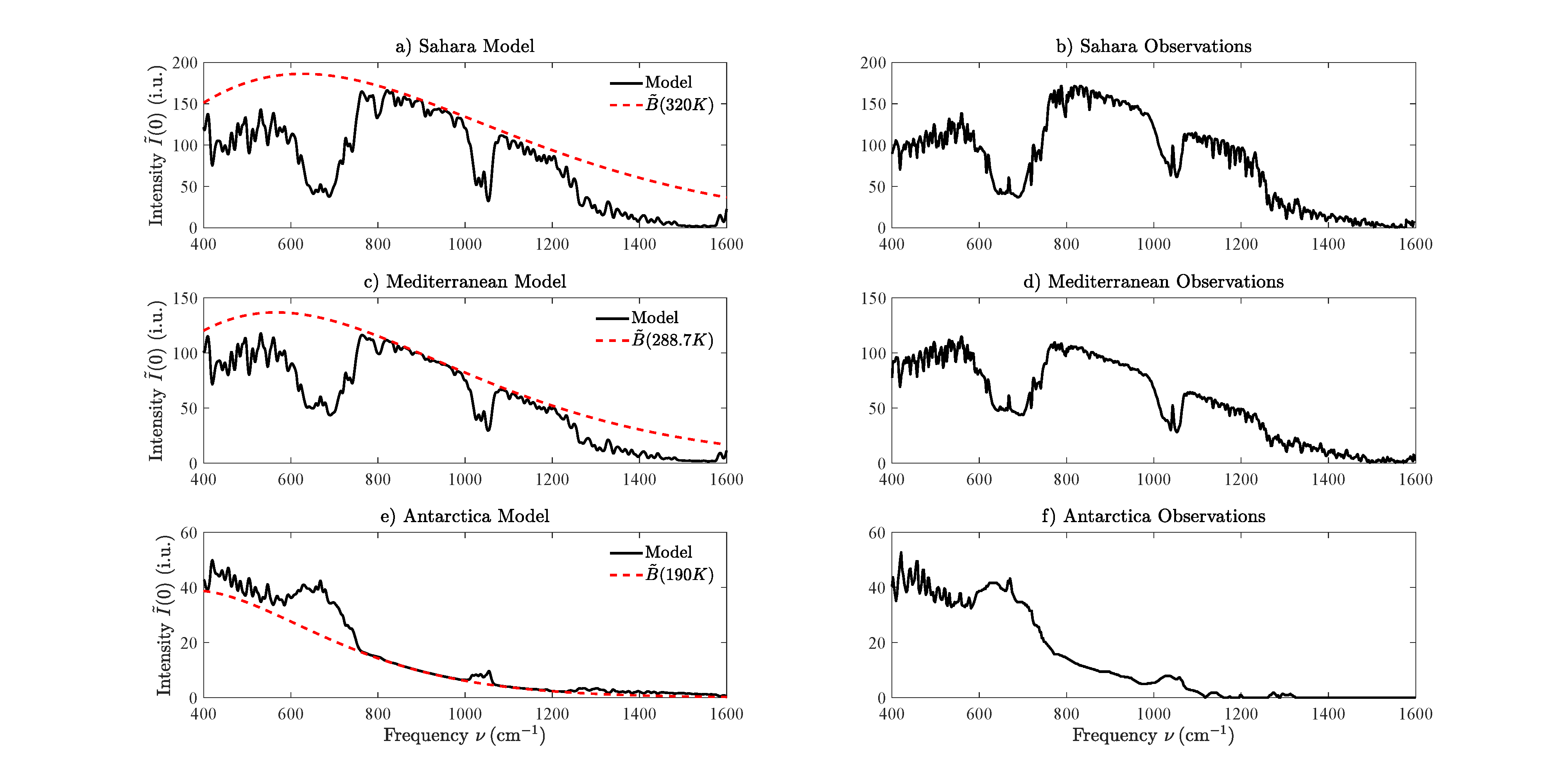} 
\caption{Comparison of modelled and satellite observed intensity of infrared radiation or heat radiated by the Earth and its atmosphere over the Sahara, Mediterranean and Antarctica \cite{vW2020}.  The intensity observed over the Sahara, whose surface temperature is a very warm 47 Celsius or 320 Kelvins, is much higher than the intensity detected over cold Antarctica where the surface temperature is only -83 Celsius or 190 Kelvins.  The dashed red curves are the intensities that would be radiated by the surface, for a transparent atmosphere without any greenhouse gases.  For Antarctica, very dense cold air sinks to the surface causing the surface temperature to be lower than the atmosphere a few kilometers above the surface.  This slightly warmer air causes the satellite to see a higher intensity than one would expect from the very cold surface.   
\label{Nimbus}}
\end{figure}

For the case of a clear sky, one can compute the amount of heat radiated by the Earth as is shown in Fig. \ref{CO2} \cite{vW2020}.  This shows the heat radiated per unit area by the Earth and its atmosphere to outer space at the various infrared colours or frequencies.  This calculation considered the 5 most important naturally occurring greenhouse gases:  water vapour (H$_2$O), carbon dioxide (CO$_2$), ozone (O$_3$), nitrous oxide (N$_2$O) and methane (CH$_4$).  The effects of water vapour are very noticeable at frequencies less than 550 cm$^{-1}$ and in the range of 1200 to 2800 cm$^{-1}$.  The two noticeable dips in the black and red curves are due to absorption by carbon dioxide at 660 cm$^{-1}$ and ozone at 1050 cm$^{-1}$.  The effects of methane and nitrous oxide are barely noticeably because their atmospheric concentrations are over 1000 times smaller than water vapour and more than 100 times smaller than carbon dioxide.  One can barely distinguish between the black and red curves which correspond to carbon dioxide concentrations of 400 and 800 ppm, respectively.  An important check of this work is to compare modelled results to data observed by satellites that measure the intensity of infrared light radiated by the Earth and its atmosphere as shown in Fig. \ref{Nimbus} \cite{vW2020}.

\subsection{Climate Sensitivity}

The effect of doubling carbon dioxide is shown by the red and black curves in Fig. \ref{CO2}.  Slightly less heat is radiated when the CO$_2$ is doubled to 800 ppm. For the Earth to be in equilibrium, the heat radiated equals the amount of absorbed sunlight.  If less heat is radiated, then the Earth warms.  The temperature increases by 1 $^o$C if the warming is the same at all altitudes.  

The surface warming $\Delta T$, due to the carbon dioxide concentration increasing from $C_o$ to $C$ is given by

\begin{equation}
\Delta T = S{\ } log_2{\ } {C \over C_o}
\label{lbl1}
\end{equation}

\begin{figure}[t] \centering
\includegraphics[height=60mm,width=.6\columnwidth]{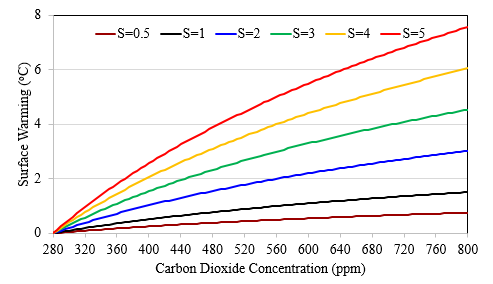} 
\caption{Surface Warming Dependence on Carbon Dioxide Concentration for various values of Climate Sensitivity $S$ in units of $^o$C as given by equation \ref{lbl1}.  The carbon dioxide concentration at the onset of the industrial revolution was 280 ppm. In 2021, the CO$_2$ concentration reached 415 ppm.
\label{ClimSens}}
\end{figure}

\noindent where $S$ is called the climate sensitivity.  The warming predicted by this formula is plotted in Fig. 5.  If the carbon dioxide concentration doubles from 400 to 800 ppm, the surface temperature increases by $S$.  The logarithm means that for warming by $2S$, the CO$_2$ concentration must double again from 800 to 1600 ppm.

\subsection{Water Feedback}

A warming of only 1 $^o$C is not a threat to the planet.  However, the maximum amount of water vapour that can be contained in a given volume of air increases at a rate of 6\% per degree Celsius.  Therefore, the water vapour concentration would go up in a warming atmosphere if the relative humidity remained unchanged.  This so called water feedback effect would amplify the small warming due to increasing carbon dioxide.  Observations are inconclusive that water vapour has increased at the maximum theoretical rate \cite{Santer}.  A study that examined 1/4 billion hourly values of temperature and relative humidity at 309 stations located throughout North America during 1948-2010 found relative humidity decreased at many inland stations \cite{VIvW2012}. 

A more sophisticated analysis of surface warming must take into account that the troposphere warms more than the higher atmosphere.  S. Manabe, who received the Nobel prize in 2021, originally estimated a climate sensitivity of 1.4 C for the case of constant water vapour concentration \cite{Manabe}.  This has been independently confirmed by several subsequent groups including our own calculations \cite{vW2020}.  The climate sensitivity increases to about 2.3 C if one considers maximum possible water feedback such that relative humidity remains constant.

\subsection{Global Warming due to Other Greenhouse Gases such as CH$_4$ and N$_2$O}

All five of the Earth's naturally occurring greenhouse gases, H$_2$O, CO$_2$, O$_3$, N$_2$O and CH$_4$ are very strongly saturated \cite{vW2020}.  This means that doubling any of their concentrations produces a global warming that is greater than a subsequent similar absolute increase.  N$_2$O and CH$_4$ are much less abundant than H$_2$O and CO$_2$ in the Earth's atmosphere and therefore somewhat less saturated.  An analogy to explain saturation is a farmer painting a barn.  One notices a big difference between the first and second coat but negligible effect between the 10th and 11th coat.  

For the case of methane, we recomputed Fig. \ref{CO2} for ambient and doubled methane concentrations of 1.8 and 3.6 ppm, respectively.   One additional methane molecule is 30 times more effective for global warming than one additional CO$_2$ molecule.  However, 300 times more carbon dioxide is added to the atmosphere each year than CH$_4$.  Hence, the overall effect of the increasing methane concentration to global warming is less than 1/10 that of carbon dioxide \cite{vW2020}.

For the case of nitrous oxide, one additional N$_2$O molecule is 230 times more effective for global warming than one additional CO$_2$ molecule,  But the rate of increase of CO$_2$ molecules is 3000 times larger than that of N$_2$O.  So the contribution of nitrous oxide to global warming is about 230/300 or about 1/13 that of carbon dioxide \cite{WH2022}.

\subsection{Global Climate Models}

Efforts to construct a realistic mathematical computer model of the Earth's climate are an enormous task that have been ongoing for over half a century \cite{GFDL}.  The Earth's surface area of 515 million km$^2$ is divided into area segments and the atmosphere into vertical slices.  For an area element of 100 km$^2$ and a vertical step size of 100 m, one needs over 2.5 billion points.  At each point, one must calculate: temperature, pressure, water vapor concentration, greenhouse gas and aerosol concentrations, wind speed and direction etc.  At the surface, one must determine  the fraction of sunlight reflected.  Evaporation must be taken into account which is very different for an ocean as opposed to a land surface.  For the latter case, the vegetation type may be important.  Ocean currents such as the Gulf Stream that transport enormous amounts of heat from the tropics to polar latitudes must also be modelled using an additional array of points.  Ocean circulation is not well understood and cycling times of deep ocean water are estimated to take up to hundreds of years.  Finally, the entire system is time dependent, as the Earth orbits annually about the sun and daily rotates about its axis.

\begin{table}
\begin{center}
\begin{tabular}{|c|c|}
\hline
IPCC Report  &Climate Sensitivity \\ [0.5ex]
Year &$S$ ($^o$C)\\
\hline\hline
1990 \cite{IPCC1990}  &1.5 - 4.5\\
\hline
1996 \cite{IPCC1996} &1.5 - 4.5\\
\hline
 2001 \cite{IPCC2001} &1.5 - 4.5\\
\hline
 2007 \cite{IPCC2007} &2.0 - 4.5\\
\hline
 2014 \cite{IPCC2014} &1.5 - 4.5\\
\hline
 2021 \cite{IPCC2021} &2.5 - 4.0\\
\hline
\end{tabular}
\end{center}
\caption{Values for Climate Sensitivity $S$, the global warming caused by doubling carbon dioxide, as given in various IPCC Reports.
\label{ClimSens}}
\end{table}

Developing a Global Climate Model (GCM) requires a very large team of people and the computer programs easily exceed a million lines of code.  GCMs have been developed at large centers such as the Princeton Gas Fluid Dyanmic Laboratory \cite{GFDL}, UK Meteorological Office \cite{Hadley}, Boulder's National Climate Center for Atmospheric Research \cite{Boulder}, Max Planck Institut f\"ur Meteorologie \cite{MPI} and the Canadian Centre for Climate Modelling and Analysis \cite{ECVictoria}.  Even using the world's fastest supercomputers, numerous simplifying approximations are essential.  Typically, one tunes a myriad of parameters to ensure the GCM reasonably reproduces the Earth's 20th century climate.  One then models future climate considering various scenarios of increasing greenhouse gases.

Work is ongoing to make GCMs more realistic.  The Coupled Model Intercomparison Project (CMIP) attempts to evaluate and compare the different GCM results \cite{CMIP}.  Table \ref{ClimSens} lists the estimates for the climate sensitivity.  These values have not changed spectacularly over 30 years.  In 2021, the IPCC gave a value for $S$ of between 2.5 and 4.0 $^o$C with the best estimate of 3 $^o$C.  The 2021 uncertainty is smaller than for previous reports because it is believed the latest GCMs better model clouds.  
It is very challenging to determine the global warming or cooling effects of clouds, which at any time cover about two thirds of the Earth.  Cooling occurs when clouds block the midday sun on a summer day while in winter, cloudy nights are warmer than clear nights.  The huge uncertainty due to clouds is the elephant in the room in Climate Science.   

\begin{figure}\centering
\includegraphics[height=70mm,width=.6\columnwidth]{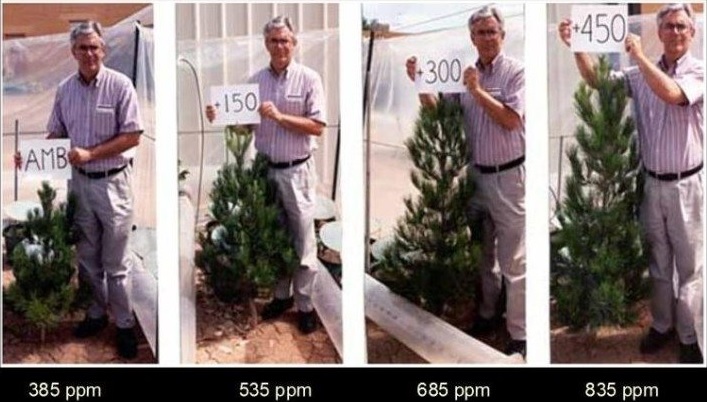}
\caption{Growth of Eldarica Pine Trees at CO$_2$ Concentrations of 385, 535, 685 and 835 ppm \cite{Idso}. 
\label{EldaricaPineTrees}}
\end{figure}

\section{Is Carbon Dioxide Beneficial to Life?}

\begin{figure}\centering
	\includegraphics[height=80mm,width=.6\columnwidth]{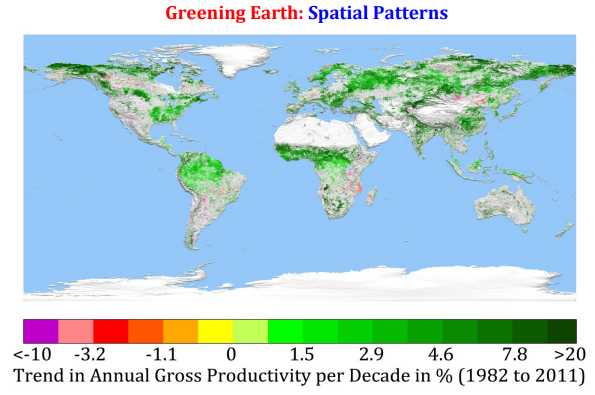}
	\caption{Global Trends in Gross Vegetative Productivity 1982-2011 from Satellite Measurements \cite{Myneni2013}.  The increase of plant growth is especially noticeable in arid regions such as south of the Sahara desert.  At higher CO$_2$ levels, the stomata in plant leaves are smaller, reducing the loss of water \cite{CO2Coalition}.  
	\label{GlobalGreening2}}
\end{figure}

Carbon dioxide is an odourless, invisible gas.  It is plant food.  Greenhouse growers routinely add it to promote growth.  An example is shown in Fig. \ref{EldaricaPineTrees} where pine trees grow much faster at higher concentrations of carbon dioxide \cite{Idso}.                                                                                                                     
An increase in global vegetative productivity has been observed by satellites as shown in Fig. \ref{GlobalGreening2} \cite{Myneni2013,Myneni2016,Modis}.  For the period 1982-2011, more than 30\% of the Earth's land area greened for a 14\% overall increase in total gross vegetative productivity.  This is very beneficial to farmers who provide food for the world's growing population \cite{Idso2011}.

Fig. \ref{GlobalGreening2} shows particularly striking greening in the region bordering the Sahara desert.  Plants extract carbon dioxide from the atmosphere through small holes in their leaves called stomata.  At higher carbon dioxide levels, the stomata become smaller, reducing water loss from the plant \cite{Idso2011,CO2Coalition}.  This is believed to be part of the reason for the notable greening of the Earth's arid regions. 

The response of plants to higher atmospheric carbon dioxide levels depends on plant type \cite{CO2Coalition}.  So called C3 plants such as soybeans and wheat, use an enzyme, rubisco, to convert CO$_2$ to simple carboyhydrate molecules having 3 carbon atoms.  These molecules are subsequently converted into more complex molecules such as sugar.  Rubisco evolved when atmospheric CO$_2$ levels were several thousand ppm rather thn 415 ppm in 2021.  At today's low CO$_2$ level, the leaf uses up much of the CO$_2$ when the sun shines.  If there is insufficient  CO$_2$, then rubisco uses an O$_2$ molecule which produces toxic bypoducts such as hydrogen peroxide \cite{CO2Coalition}

C4 plants such as corn evolved during the past tens of millions of years when atmospheric CO$_2$ levels were relatively low.  These plants developed special structures within the leaf to prevent the reaction of rubisco with oxygen.  CO$_2$ molecules are ferried into the protective leaf structure by molecules having 4 carbon atoms.  It is reasonable to expect that C3 and C4 plants would respond differently to increased CO$_2$ levels.
 
There has been a huge increase in agricultural productivity in the last century.  American soybean and corn yields since 1940 have increased three and six fold respectively \cite {Taylor}.  Most of this increase is due to better farming techniques including improved genetics, irrigation and the use of chemical fertilizers.  The effects of increased atmospheric CO$_2$ however are not negligible.  A recent paper states ``a 1 ppm increase in CO$_2$ equates to a 0.5\%, 0.6\% and 0.8\% yield increase for corn, soybeans and wheat respectively.  Viewed retrospectively, 10\%, 30\% and 40\% of each crop's yield improvements since 1940 are attributable to rising CO$_2$" \cite{Taylor}. Other studies have also found higher atmospheric CO$_2$ levels increase plant growth.  For example, a review by Poorter and Navas showed doubling CO$_2$ increases growth of C3 plants by an average of 45\% while the response of C4 plants is about 12\% \cite{Poorter}.
  
\section{Are Greenhouse Gas Concentrations Increasing?}

\begin{figure}\centering
	\includegraphics[height=100mm,width=.5\columnwidth]{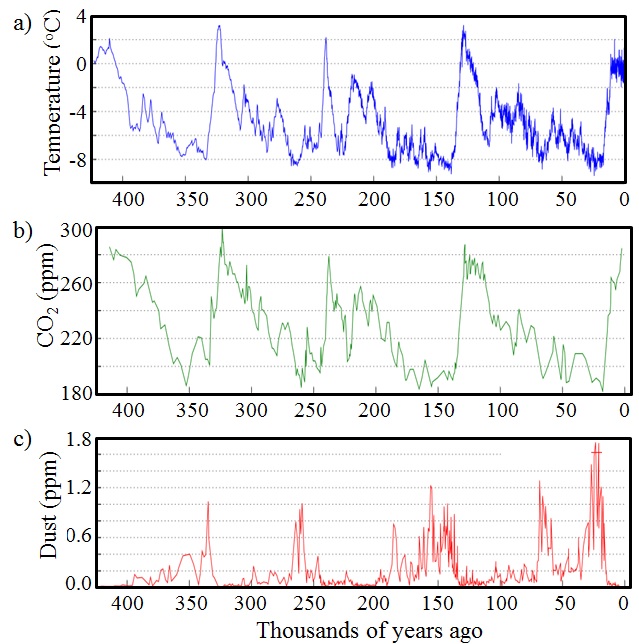}
	\caption{Vostok, Antarctica Ice Core Determination of a) Temperature Change relative to the present b) CO$_2$ Concentration and c) Dust Concentration for the last 400,000 years \cite {Vostok}.  Temperature and the atmospheric CO$_2$ concentration are strongly correlated.  Careful examination has shown that the temperature increases shown in (a) precede the increases in the CO$_2$ level given in (b) by about 800 years \cite{Caillon}.
		\label{Vostok}}
\end{figure}

Atmospheric carbon dioxide levels have varied throughout Earth's history.  They were as high as 4000 ppm 500 million years ago in the Cambrian period and dropped to 180 ppm at the peak of the last ice age \cite{Berner2001}.  Carbon dioxide concentrations of the last several hundred thousand years are shown in Fig. \ref{Vostok}.  This data was obtained from ice cores extracted at Vostok, Antarctica \cite{Vostok}.  Each year a layer of snow as well as dust falls onto the surface.  The weight of succeeding layers compress the snow which eventually turns to ice.  Air is then trapped in small bubbles.  Fig. \ref{Vostok} shows the CO$_2$ concentration measured as a function of the depth of ice retrieved from a borehole at Vostok.  The age of the ice is determined by counting the annually deposited ice layers analogous to counting tree rings.  The temperature is found by measuring the abundance of the heavier isotopes of oxygen and hydrogen in the ice \cite{Riebeek}.  Regular oxygen has 8 protons and 8 neutrons and is labelled $^{16}$O.  A few oxygen atoms have 2 additional neutrons and are labelled $^{18}$O.  Similarly, some hydrogen atoms have an additional neutron and are called deuterium.  These heavier isotopes cause some water molecules to be slightly heavier.  It has been observed that the concentration of these heavy water molecules in snow depends on the temperature.  The reason is that more energy is required to evaporate a heavier water molecule.  Hence, at colder temperatures the abundance of these isotopes in water vapour is less which in turn lowers their concentration in snow.  This allows the temperature to be determined.

Fig. \ref{Vostok} shows a very strong correlation between temperature change and carbon dixoide concentration although less so with dust that is produced by volcanoes.  The obvious question is whether a carbon dioxide increase precedes a temperature increase or vice versa.  A number of careful examinations have shown that the increase in CO$_2$ occurred about 800 years after temperature began to increase \cite{Caillon}.  The increase in CO$_2$ is believed to have been caused by a warming ocean releasing carbon dioxide.  A similar effect happens when a bottle of coca cola is warmed producing bubbles of carbon dioxide.  Hence, increasing atmospheric carbon dioxide was not responsible for the initial temperature rise.  The question remains as to what triggered the temperature increase and why it decreased at a later date when the CO$_2$ concentration was elevated.  

\begin{figure}\centering
\includegraphics[height=140mm,width=0.5\columnwidth]{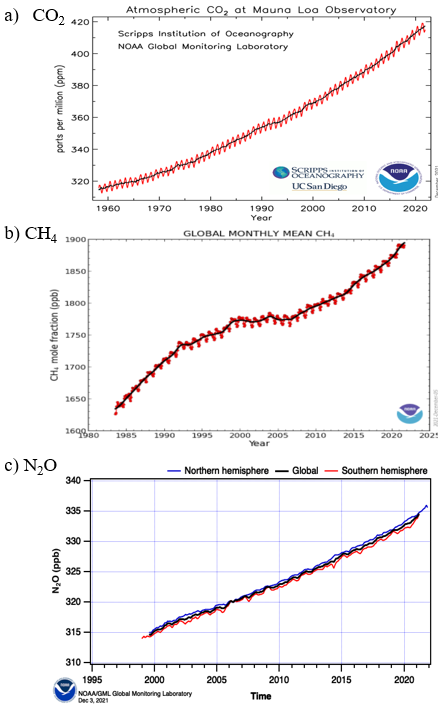}
\caption{Observations made by NOAA's Global Monitoring Laboratory at Mauna Loa of a) CO$_2$, b) CH$_4$ and c) N$_2$O \cite{MaunaLoa}.  The increases in CO$_2$ and CH$_4$ are primarily due to the combustion of fossil fuels while the increase in N$_2$O is caused by greater use of nitrogen based fertilizers.  The annual oscillation of CO$_2$ and CH$_4$ are caused by the seasonal dependence of biological activity.  More of the Earth's land mass is located in the Northern as opposed to the Southern hemisphere.  Plants absorb CO$_2$ during spring and summer.  It is released back to the atmosphere in the fall and winter when the vegetation decays.  The rate of CH$_4$ increase fell sharply in the late 1990s and up to about 2010.  A possible reason for this is that leaking Russian Natural Gas pipelines were fixed.   
\label{MaunaLoa}}
\end{figure}

Fig. \ref{MaunaLoa} shows more recently measured concentrations of the greenhouse gases: CO$_2$, CH$_4$ and N$_2$O made at Mauna Loa, Hawaii.  This location is distant from large industrialized cities and therefore more likely to accurately reflect the average global levels of these gases.  Since 1958, the CO$_2$ concentration has risen from 315 to 415 parts per million (ppm) in 2021.  The 2020 rate of CO$_2$ increase is about 0.5\%/year or about 2.5 ppm/year.  The preindustrial concentrations of CO$_2$, CH$_4$ and N$_2$O are estimated to be 280, 0.7 and 0.26 ppm, respectively.  The gas concentrations vary throughout the year.  The CO$_2$ level decreases in spring and summer and increases during fall and winter by about 6 ppm.  This is caused by biological activity.  Most of the Earth's land mass is located in the Northern Hemisphere.  Carbon dioxide is absorbed by plants during spring and summer.  Correspondingly, vegetation decays in the fall and winter releasing carbon dioxide as well as methane back into the atmosphere.

Fig. \ref{MaunaLoa}b shows methane concentrations stopped increasing during the period from about 1998 to 2008.  The CH$_4$ concentration actually decreased in a few of those years!  The reason for this decline is not entirely clear.  It has been noted that the Russian Natural Gas industry replaced old leaking pipelines during this period.  Recently, there have been reports of massive leaks from Russian pipelines during the coronavirus epidemic due to a lack of maintenance \cite{Gerretsen}. 

The increase of carbon dioxide is due primarily to the combustion of fossil fuels such as oil, coal and natural gas that resulted from the industrial revolution that began in about 1750. There are significant other sources of greenhouse gases.  As much as 20\% of the annual anthropogenic production of carbon dioxide has been estimated to result from clearing forests by burning vegetation \cite{Gullison,vdW}.  The most notable deforestation has occurred in Brazil where about 10\% or 400,000 km$^2$ of the Amazonian rainforest was destroyed during 1988-2013 \cite{Brazil}.

\begin{figure}\centering
\includegraphics[height=80mm,width=.6\columnwidth]{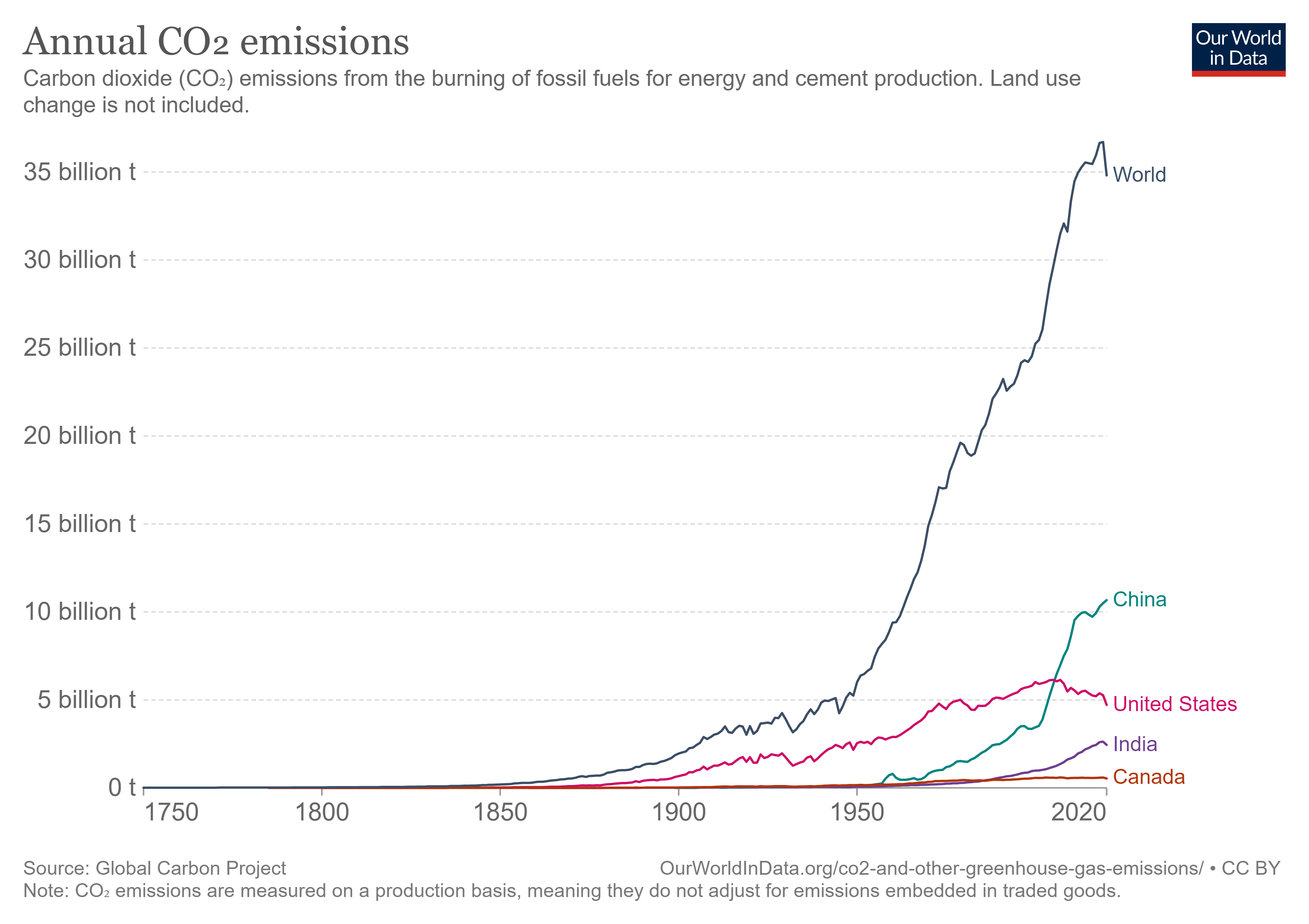}
\caption{Annual emissions of CO$_2$ measured in billion metric tons from the burning of fossil fuels for energy and cement production for the World, China, United States, India and Canada \cite{AnnualCO2}.   
\label{AnnualCO2}}
\end{figure}

\begin{figure}\centering
	\includegraphics[height=80mm,width=.8\columnwidth]{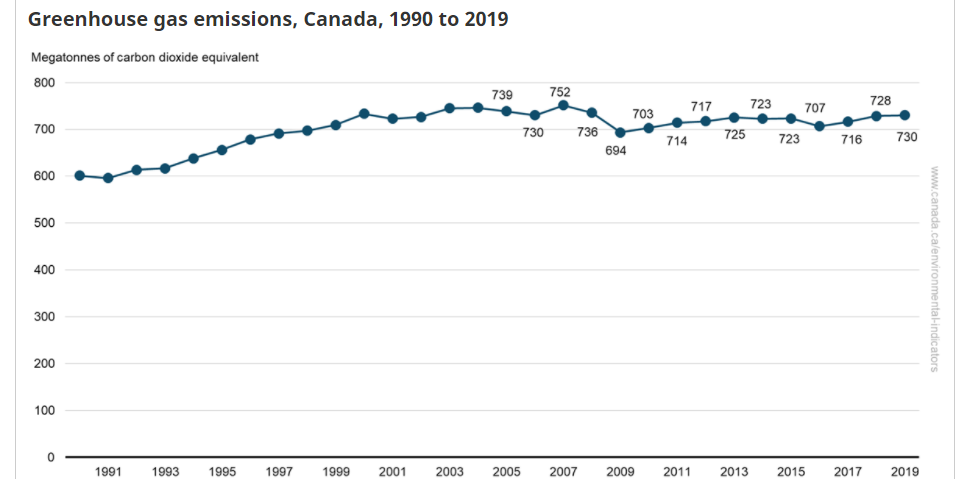}
	\caption{Megatonnes of CO$_2$ Equivalent Emissions by Canada from 1990 to 2019 as given by the Environment and Natural Resources Department of the Government of Canada \cite{CanadaProvEmissions}. 
		\label{CanadaEmissions}}
\end{figure}

\begin{figure}\centering
	\includegraphics[height=80mm,width=.8\columnwidth]{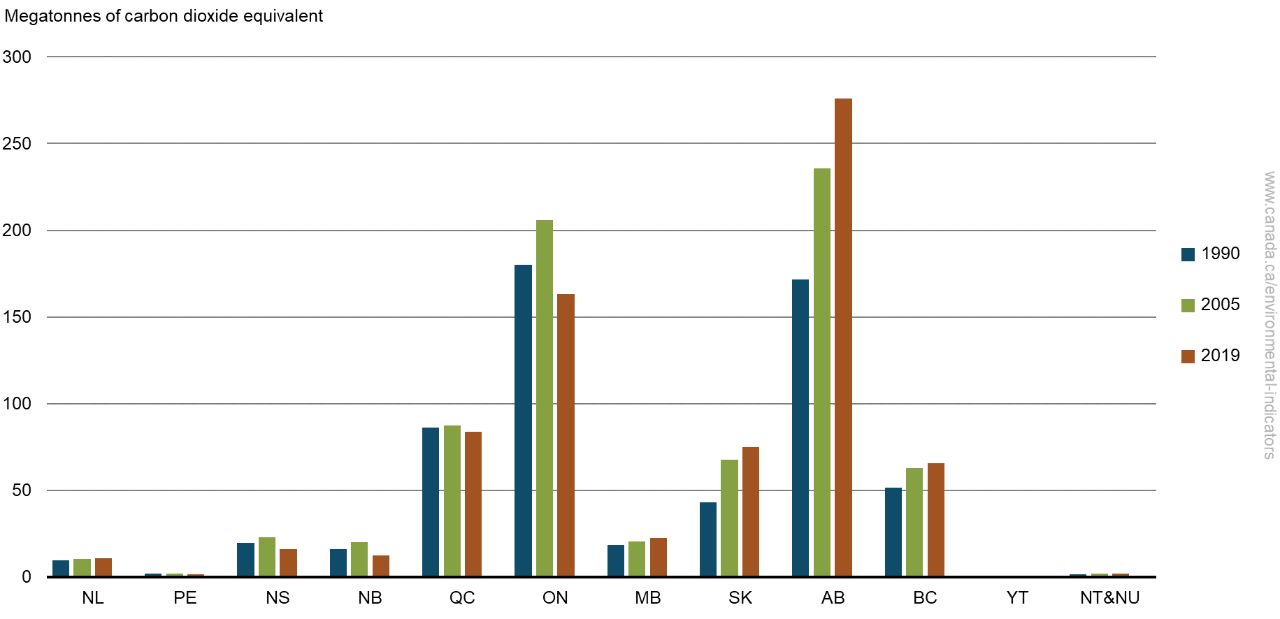}
	\caption{Megatonnes of CO$_2$ Equivalent Emissions by Province in Canada for 1990, 2005 and 2019 as given by the Environment and Natural Resources Department of the Government of Canada \cite{CanadaProvEmissions}.  The 20\% decrease in Ontario's CO$_2$ equivalent emissions from 2005 to 2019 resulted from closing coal burning electric generating plants.
		\label{CanadaProvEmissions}}
\end{figure}

Fig. \ref{AnnualCO2} shows Canada produced 582.4 megatonnes of CO$_2$ from burning fossil fuels for energy and cement production in 2019 \cite{AnnualCO2}.  This represents 1.6\% of the world total of 36.7 billion tons.  
The largest emitters are China, United States and India.  The amount of CO$_2$ generated by the United States has decreased over the last 5 years as natural gas increasingly replaces coal.  Both natural gas and coal produce CO$_2$ when burned but natural gas produces more energy per amount of CO$_2$.  The amount of CO$_2$ produced by China and India has been accelerating as these countries develop their economies to raise living standards.  The slight decrease in CO$_2$ generated by the world in 2020 is due to decreased economic activity resulting from the corona virus pandemic.  

A detailed look at Canada's emission from 1990 to 2019 is given in Fig. \ref{CanadaEmissions}.  This plots megatonnes of carbon dioxide equivalent which is greater than the amount of CO$_2$ produced because it includes the estimated global warming effect of other greenhouse gases, notably CH$_4$ and N$_2$O \cite{CanadaProvEmissions}.  Fig. \ref{CanadaProvEmissions} shows the megatonnes of carbon dioxide equivalent produced in 1990, 2005 and 2019 by each of Canada's provinces \cite{CanadaProvEmissions}.  
In 2019, Ontario produced 163.2 megatonnes equivalent of CO$_2$ or 22\% of Canada's total of 730 megatonnes.
Ontario was the only province to show a sizable decrease from 205.7 megatonnes produced in 2005.  This decrease resulted from its closure of coal burning electric generating plants.
In the 2 years prior to the corona virus pandemic, global CO$_2$ production increased from 35.93 billion tons in 2017 to 36.70 billion tons in 2019 corresponding to an average increase of 385 megatonnes per year.  This is more than double Ontario's total annual CO$_2$ production.  If Ontario were to suddenly stop using all fossil fuels, it would barely be noticed on the global scale.  Note that stopping all fossil fuel use means no gasoline for motor vehicles, no natural gas to heat homes during winter nor produce electricity to power electric cars etc. 

\begin{table}
	\begin{center}
		\begin{tabular}{|c|c|c c|}
			\hline
			&&&\\
			Region &{CO$_2$ Increase (ppm/year)}  
			&\multicolumn{2}{c|}{Warming ($^o$C/year)}\\	
			&  &$S=1$ $^o$C &$S=3$ $^o$C \\ [0.5ex]
			\hline\hline
			&&&\\
			World   & 2.5  &0.0087  &0.026\\
			&&&\\
			\hline
			&&&\\
			Canada  &1.6\% $\times$ 2.5 = 0.040 &$1.4 \times 10^{-4}$  &$4.2 \times 10^{-4}$\\
			&&&\\
			\hline
			&&&\\
			Ontario  &22\% $\times$ 0.040 = 0.0088 &$3.1 \times 10^{-5}$  &$9.2 \times 10^{-5}$\\
			&&&\\			
			\hline
		\end{tabular}
	\end{center}
	
\caption{Contribution to Global Warming by Canada and Ontario.  Here, we consider CO$_2$ increasing from its 2021 concentration of 415 ppm at the current observed rate of 0.5\% per year or 2.5 ppm/year for climate sensitivities $S=1$ $^o$C and $S=3$ $^o$C. The warming is computed using equation (1).  The 2021 IPCC report gives the most likely value of the climate sensitivity to be $S=3$ $^o$C \cite{IPCC2021}. \label{CanOntGlobalWarming}}
\end{table}

The contribution of Ontario to global warming is given in Table \ref{CanOntGlobalWarming} for climate sensitivity values $S=1$ and $S=3$ $^o$C, the value recommended by the 2021 IPCC.  Ontario's contribution is estimated by multiplying Ontario's 2019 22\% fraction of Canada's CO$_2$ production by 1.6\%, Canada's fraction of global CO$_2$ emission, giving 0.35\% of the world total CO$_2$ emission of 2.5 ppm/year.  Using equation (1) one finds Ontario is responsible for $3.1 \times 10^{-5}$ $S$ $^o$C/year or 31 $S$ microdegree C/year (micro = one millionth) rise in global temperature.  It should be emphasized that this maximum estimate was obtained using the drastic scenario of an immediate complete cessation of all fossil fuel use.  If one were to aim at a more palatable 5\% annual reduction, the corresponding effect on global climate would be twenty times smaller.  

\section{Comparison of Observations to Global Climate Model Predictions}
\subsection{Temperature}

\begin{figure}\centering
\includegraphics[height=80mm,width=.6\columnwidth]{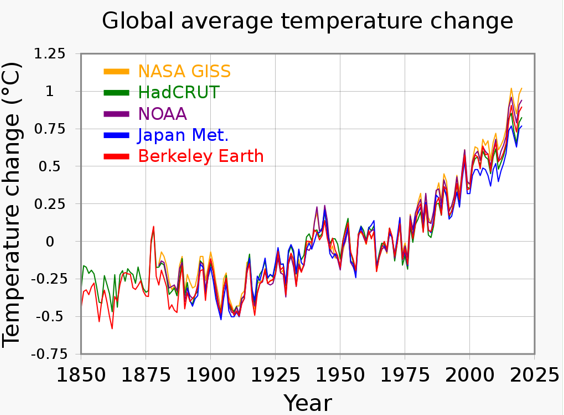}
\caption{Change of Earth's Average Surface Temperature relative to that during 1951-1980 \cite{TempObs}.  The temperature increased by about 0.5 $^o$C from about 1900 to 1940.  It decreased slightly during 1940 to 1980.  Temperature rose by about 0.5 $^o$ from 1980 to 2000.  The Earth's average temperature remained relatively unchanged from 2000 to 2016 which is known as the global warming hiatus.  Global climate models have difficulty explaining this decadal temperature variation and failed to predict the hiatus after 2000.
\label{TempObs}}
\end{figure}

Fig. \ref{TempObs} shows the observed change of the average Earth's surface temperature over the last century and a half. The temperature has increased by about 1 $^o$C over this time.  The curve shows a temperature increase as one would expect from Figures 5 and \ref{MaunaLoa}.  However, there are notable differences. 

1.  From about 1900 to 1940, there was a warming of about 0.5 $^o$C during which CO$_2$ increased from 296 to 311 ppm \cite{CO2Level}.  Using equation (1), this gives a climate sensitivity of $S=7$ which greatly exceeds the values given in Table \ref{ClimSens}.  Therefore, most of this 0.5 $^o$C warming is not believed to be due to increasing greenhouse gases but due to natural variation or continued warming of the Earth following the end of the Little Ice Age. 

2.  From 1940 to 1980, there was a slight temperature decrease.  The reason for this is not clear, especially since the CO$_2$ concentration increased from 311 to 339 ppm.  

3.  From 1980 to 2000, temperature increased by about 0.5 $^o$C while CO$_2$ increased from 339 to 370 ppm.  

4.  From 2000 to about 2016, the temperature remained stable which is known as the global warming hiatus.  This was completely unexpected as the CO$_2$ concentration increased from 370 to 404 ppm.  

\begin{figure}\centering
	\includegraphics[height=150mm,width=.7\columnwidth]{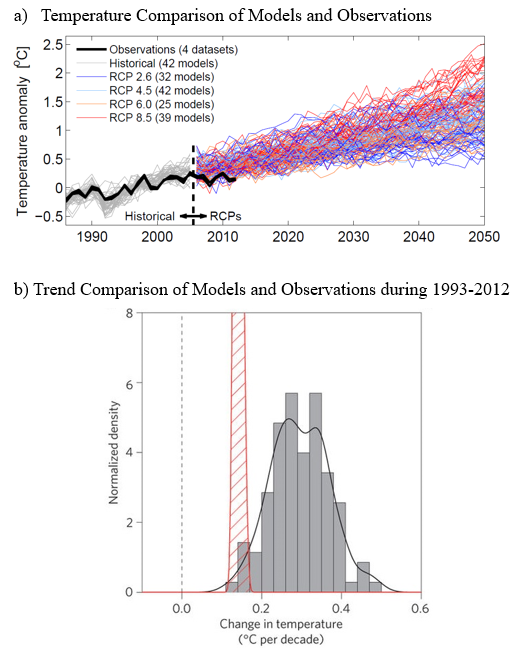}
	\caption{Comparison of Global Climate Model Projected \cite{IPCC2014} and Observed a) Temperature and b) Temperature Trends during 1993 to 2012 \cite{Fyfe}.  Nearly all climate models (grey) significantly exceed the observed (cross hatched red) global warming.
		\label{ModelObsComp}}
\end{figure}

The failure of the GCMs to predict the hiatus in global average temperature after 2000 is illustrated in Fig. \ref{ModelObsComp}. The observed temperature increase per decade during 1993 to 2012 is less than half predicted by nearly all GCMs. 

The excessive warming projections continues to be a problem for the latest  IPCC 2021 report.  A study compared the observed average surface temperature increase between 1980 and 2021 of 0.56 $^o$C to that given by the 38 GCMs \cite{Scafetta}.  Every GCM having a climate sensitivity greater than 3.0 $^o$C predicted a higher temperature increase than was observed  Only the GCMs having the lowest sensitivity between 1.8 and 3.0 $^o$C agree with observations.  This indicates that the 2021 IPCC's estimate of 3 $^o$C for the climate sensitivity is likely too high. 

\begin{figure}\centering
	\includegraphics[height=150mm,width=.6\columnwidth]{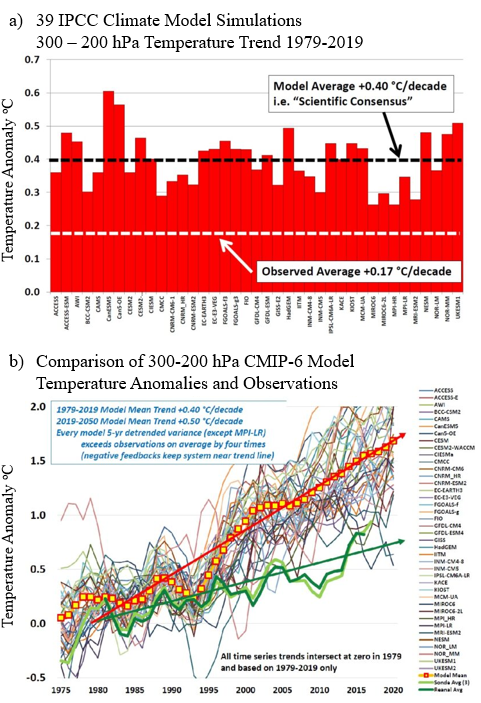}
	\caption{Comparison of Climate Model Simulated and Observed Temperatures at altitude of 10 - 13 km. a) Trend Comparison for 1979-2019 and b) Projected Temperature Anomaly and Observations (green). The average of the climate model runs is shown by the yellow dots bordered with red \cite{Christy2021}.  The observed temperature is substantially below values predicted by nearly all GCMs.
		\label{CMIP6TropTemp}}
\end{figure}

\begin{figure}\centering
	\includegraphics[height=80mm,width=.8\columnwidth]{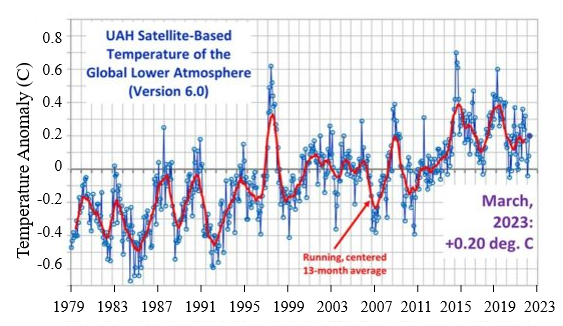}
	\caption{Average temperatures from 1979 to 2023 at the tropopause altitude as measured by NOAA satellites that detect microwave emissions from oxygen \cite{Spencer2023}.  The temperature anomaly is the difference of the temperature and the average observed from 1991 to 2020.  The observed trend is about 0.15 C per decade consistent with the data shown in Fig. 15 for the period 1979-2019.
		\label{Spencer2023Edit}}
\end{figure}

The largest global warming is expected to occur at the tropopause, the boundary of the troposphere and stratosphere \cite{vW2020}.  For midlatitudes, the tropopause occurs at an altitude of 11 km where atmospheric pressure is about 300 to 200 hPa.  Atmospheric pressure at the Earth's surface is just over 1,000 hPa. Fig. \ref{CMIP6TropTemp} compares the results of the latest Coupled Model Intercomparison Project (CMIP6) to observations.  It should be emphasized that these were the climate models used in the latest IPCC 2021 report.  The modelled temperature trend for 1979-2019 is more than twice what was observed.  The observed temperature anomaly, given by the green curves in Fig. \ref{CMIP6TropTemp}b as well as in Fig. \ref{Spencer2023Edit} for the period 1979-2023, is also much less than predicted by nearly every GCM.  This shows GCMs systematically overestimate global warming \cite{Fyfe}.

The observed data do show the Earth's average temperature has warmed since 1880.  Warming trends are especially pronounced if one compares temperatures to the period 1950 to 1980 when the average temperature did not increase. 

The warming of the Earth is not evenly distributed but is somewhat larger in the Arctic \cite{vW2015b,vW2015}.  This is not surprising given that ice reflects more sunlight than water.  However, a reduction in summer Arctic ice of 2 million km$^2$ as will be discussed in the next Section, represents less than 0.4\% of the Earth's surface and therefore has primarily a regional as opposed to a global impact.

For Ontario, global warming since 1980 is reflected in a slightly longer growing season and undoubtedly a slightly increased incidence of heat waves.  The latter should be most pronounced in large urban areas.  The contribution of increased greenhouse gases to heat waves is difficult to study as metropolitan regions, such as the Golden Horseshoe, have expanded enormously in recent decades replacing pristine meadows and farmland with asphalt, buildings etc. that heat up much more in the summer sun.  This so called urban heat island effect can increase temperatures by as much as 1 $^o$C \cite{Jones}.

\subsection{Polar Ice Caps}
The world's largest glaciers are located at the North and South Poles.  The largest volume of Arctic ice is contained in the Greenland ice sheet.  It has a volume of 2.6 million km$^3$ and covers an area of 1.7 million km$^2$ with a thickness ranging from 1 to 3 kilometers \cite{NSIDC}.  The remaining Arctic ice is floating.  So called multiyear ice has a typical thickness between 3 to 4 meters while single year ice is only about 1 meter thick \cite{Kwok}.  

\begin{figure}\centering
	\includegraphics[height=120mm,width=.6\columnwidth]{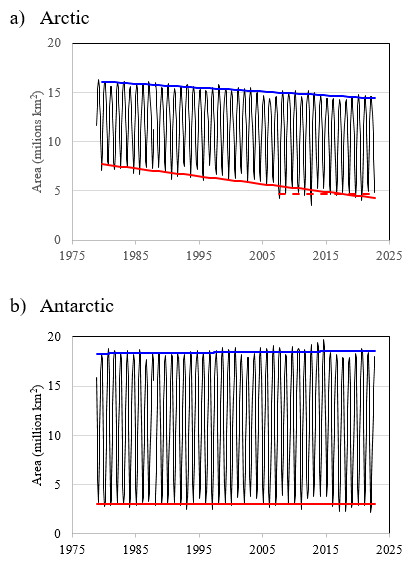}
	\caption{Sea Ice Extent (area of ocean with at least 15\% sea ice) as given by National Snow \& Ice Data Center during 1978 to 2022 for a) Arctic and b) Antarctic \cite{ArcticAntarcticIce}.   For the Arctic, the slope of the solid blue (March) and red (September) trendlines are -3.9 and -7.9 million km$^2$ per century, respectively.  For the period 2007 to 2022, the slope of the dashed red (September) trendline is -0.10 million km$^2$ per century.  Extrapolating the solid (dashed) red line leads to a prediction of an ice free Arctic ocean in 2077 (4729).  For the Antarctic, the slopes of the solid blue (September) and red (February) trendlines are +0.61 and -0.13 million km$^2$ per century, respectively.  
		\label{ArcticAntarcticIce}}
\end{figure}

The extent of the polar ice caps has been monitored since 1979 by satellites.  Fig. \ref{ArcticAntarcticIce} shows the Arctic ice cap has been steadily shrinking since 1979.  The reduction of the Arctic ice cap has been especially noticeable north of Siberia and in Greenland in recent years \cite{NSIDC}.  This is not surprising given that Arctic temperatures have increased during that period \cite{vW2015b,vW2015}.  The minimum sea ice extent has been decreasing faster than the maximum sea ice extent.  This can be explained by a decrease in the ice thickness which has also been observed by analyzing data recorded by submarines \cite{Kwok}.  A reduction of ice thickness may in part be due to storms that have dislodged multiyear ice in recent decades \cite{Lukovich}.
Extrapolation of the September red solid trendline, leads to a prediction of no ice in September, 2077.  If however, one extrapolates the dashed red trendline obtained using data for 2007 to 2022, an ice free Arctic must wait until 4729. 
These ice free projection dates can change significantly as one obtains additional years of years.  For example, the trend line obtained using 1979-2021 (2007-2021) data predicts an ice free Arctic in September 2073 (2585).  

\begin{figure}\centering
	\includegraphics[height=100mm,width=.5\columnwidth]{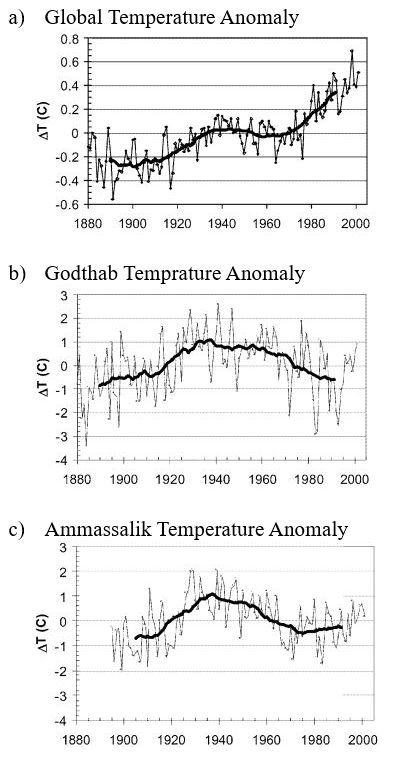}
	\caption{Temperature Anomaly for a) Global b) Godthab and c) Ammassalik in Greenland.  The solid curves are the running 21 year running averages.  The temperature anomaly is found by subtracting the average temperature \cite{Chylek}. Note the vertical scales of the temperature anomaly for Godthab and Ammassalik are larger than for the global temperature anomaly indicating that Greenland temperatures especially around 1940 were about 2 $^o$C warmer than in 2000. 
		\label{GreenlandTemperature}}
\end{figure}

In recent years, there have been a number of reports of dramatic summer melting of the Greenland icecap \cite {NSIDCGreenland}.  Over a century of data exists for a few stations in Greenland and is shown in Fig. \ref{GreenlandTemperature}.  This compares the temperature anomaly, obtained by subtracting the average temperature from the observed temperature, for the globe and two Greenland stations, Godthab and Ammassalik.
Both stations exhibit temperatures in the 1940s that were about 2 $^o$C warmer than in 2000.  This temperature anomaly is larger than the global temperature anomaly of about 0.7 $^o$C that occurred between 1965 and 2000.  The decadal variation of the Greenland temperature may be due to the so called Atlantic Multidecadal Oscillation  associated with variation of North Atlantic sea surface temperatures \cite{NAO}.  

Scientific investigations have been carried out in the Arctic for about a century.  An interesting article describes a Norwegian expedition to Spitzbergen \cite{Ifft}.  It states:  {\it ``The Arctic seems to be warming up.  Reports from fishermen, seal hunters and explorers who sail the seas about Spitzbergen and the eastern Arctic all point to a radical change in climatic conditions, and hitherto unheard of high temperatures''}.  It adds: {\it ``where formerly great masses of ice were found there are now often moraines, accumulations of earth and stones.  At many points where glaciers formerly extended far into the sea, they have entirely disappeared. ... Great shoals of white fish have disappeared and seals are few in number"}.  The waters formerly {\it ``held an even summer temperature of about 3 $^o$C; this year recorded temperatures up to 15 $^o$C and last winter the ocean did not freeze over even on the north coast of Spitzbergen"}.  The article was published in 1922.

\begin{figure}\centering
	\includegraphics[height=120mm,width=.75\columnwidth]{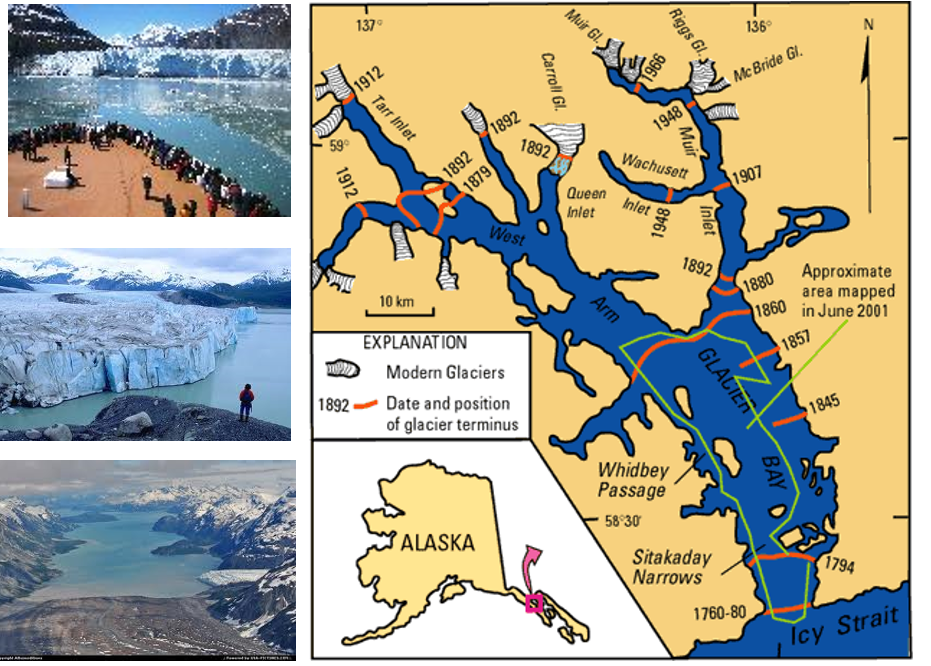}
	\caption{Glacier Bay, Alaska offers some of the most stunning images of a glacier \cite{GlacierBay}.  The red lines on the map at right show the ice ocean boundary.  The glacier has retreated dramatically over the past 200 years.
		\label{GlacierBay}}
\end{figure}

A very famous glacier is located in Glacier Bay Alaska, at a latitude just below the Arctic circle.  This glacier has retreated dramatically over the last two centuries as shown in Fig. \ref{GlacierBay}.  The glacier extended nearly to the Pacific Ocean when the region was first surveyed by George Vancouver in 1794.  Greenhouse gas emissions were negligible during the 1800s.  This indicates attributing melting glaciers as completely due to increasing greenhouse gas concentrations is incorrect.  The glacier retreat has likely been due to ongoing warming of the Earth following the end of the Little Ice Age.  

The Antarctic ice sheets are the world's largest with a volume of about 27 million km$^3$ \cite{Fretwell}.  The Antarctic ice sheet is nearly all on land although part of it is below sea level.  The ice thickness averages several kilometers with a maximum thickness of just under 5 kilometers. Fig. \ref{ArcticAntarcticIce} shows the maximum September Antarctic ice sheet has gotten slightly larger during 1979 to 2022.  The September 2014 ice area was the largest ever recorded since observations began.  The trendline for the minimum ice extent in February for 1979-2022 is nearly flat with a slope of -0.13 million km$^2$ per century.  This trend changes to +0.15 million km$^2$ per century using only the 1979-2021 data.  The ice sheet appears to be increasing everywhere around the continent with the exception of the Antarctic peninsula that juts northward toward South America \cite{Katzutoshi}.

\subsection{Oceans}
The obvious result of melting glaciers would be to increase sea level.  Various IPCC scenarios predict increases as high as 2 meters by 2100.  An increase of about 6 meters would occur if the entire Greenland ice sheet melted while the demise of the Antarctic ice sheets would increase ocean levels by over 60 meters \cite{NSIDC}.  This would inundate coastal areas displacing hundreds of millions of people.  It is important to realize that sea levels only go up if glaciers located on land melt.  The melting of the entire floating Arctic ice cap would have no effect.  The reason is that floating ice displaces a volume of water that nearly equals the volume it would occupy if the ice were converted to water. 

\begin{figure}\centering
\includegraphics[height=130mm,width=.6\columnwidth]{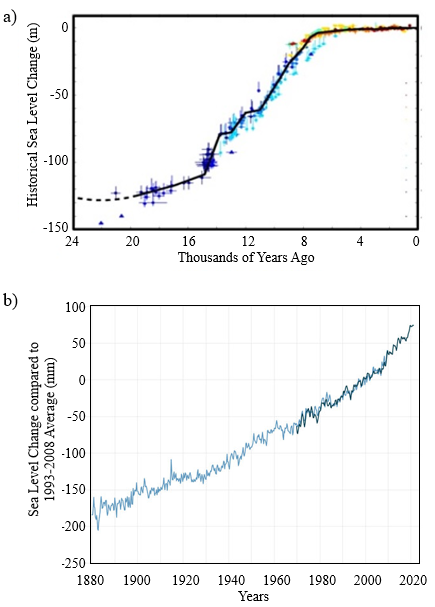}
\caption{a) Sea Level Change since the Last Ice Age \cite{PostGlacialSeaLevel} and b) Observed Sea Level Change from 1880 to 2020 \cite{Lindsey2020}.  Note the vertical scale in (a) is in meters while that in (b) is in mm.  It appears the rate of sea level rise in the 20th century of 2 mm/year increased to about 3.5 mm/year in the late 20th and early 21st centuries. 
\label{SeaLevel}}
\end{figure}

Fig. \ref{SeaLevel} shows sea level has risen over 100 meters since the end of the last ice age.  The rate of sea level rise 10,000 years ago exceeded 10 mm/year.  Ancient sea levels can be estimated using fossil coral reefs.  Coral reefs grow vertically as sea level increases such that the distance to the ocean surface remains approximately constant.  The age of successive levels of coral can be determined by measuring the concentration of radioactive isotopes such as $^{14}$C \cite{Mesollela, Eisenhauer}.

\begin{figure}\centering
\includegraphics[height=80mm,width=.6\columnwidth]{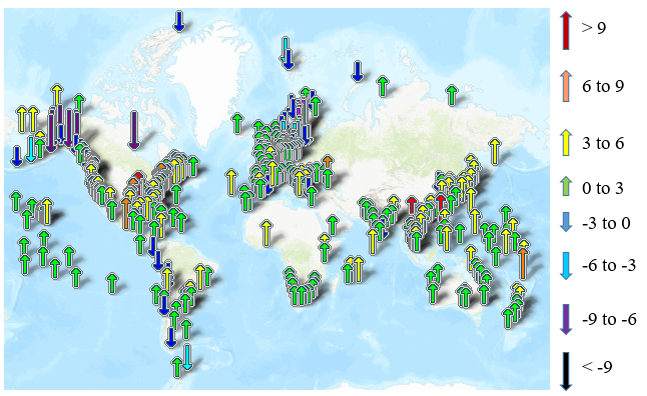}
\caption{Sea Level Trends in units of mm/year observed at stations around the world as given by the National Oceanographic and Atmospheric Administration (NOAA) \cite{GlobalSeaLevel}.  Sea level along Ontario's sea coast in Hudson Bay is observed to decrease over 9 mm/year as a result of isostatic rebounding of the land in response to the end of the last Ice Age.   
\label{SeaLevelMap}}
\end{figure}

Modern sea level measurements are determined by averaging over low and high tides.  Fig. \ref{SeaLevel} shows that sea level rose at an average rate of 2 mm/year between 1880 and 2000 \cite{Douglas}.  This increased to 3.5 mm/year for data covering the period 1994 to 2008.  Extrapolating the rate of 3.5 mm/year from 2021 to 2100 results in an average global sea level rise of 28 cm.  A significant complication in this determination is that land may be subsiding or rebounding.  The region bordering Hudson Bay is rising at about 10 mm/year \cite{Sella}.  This so called isostatic rebound occurs because the land no longer is depressed by the weight of kilometer thick ice sheets that existed during the last ice age.

The increase in sea levels are caused not just by additional water supplied by melting glaciers but also by the thermal expansion of water.  The maximum density of water occurs at 4 $^o$C.  This is approximately the ocean temperature at depths below 1000 meters.  Most of the ocean's heat is contained in the top few hundred meters.  A 200 meter column of water whose temperature changes from 20 to 21 $^o$C would increase in height by 14 mm.  Sea level should increase as the ocean absorbs heat from a warming atmosphere.  Estimates vary but it appears that this so called thermosteric contribution is responsible for a sea level increase throughout the 20th century of about 1 mm/year \cite {Lombard}.  This effect is difficult to estimate, in part because the transfer of heat via ocean currents is not well understood.  It may take decades or even centuries before heat is transferred to the deep ocean depths \cite{Marcelja}.

Ontario's contribution to global sea level rise is found by multiplying the fraction of global CO$_2$ emitted by the province, 0.35\%, by the anthropogenic rate of sea level rise.  The latter can be estimated assuming it to equal the entire change in the rate of sea level rise observed during 1994 to 2008, 3.5 mm/year relative to the rest of the 20th century, 2.0 mm/year.  One finds Ontario is responsible for 0.005 mm/year sea level rise.  This is minuscule when one compares to tides whose amplitudes are as high as 16,000 mm in the Bay of Fundy and occur about twice each day.

\begin{figure}\centering
\includegraphics[height=60mm,width=.5\columnwidth]{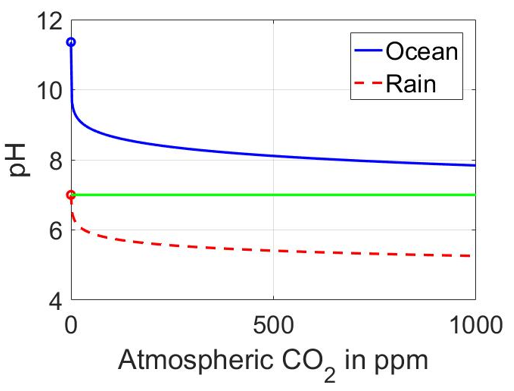}
\caption{Dependence of Ocean and Rain pH on Atmospheric CO$_2$ Concentration \cite{Cohen}.  The green line denotes neutral distilled water having a pH of 7.  It is the presence of dissolved CO$_2$ that makes the oceans habitable.  Doubling atmospheric CO$_2$ from 400 to 800 ppm would decrease ocean pH by 0.27.  
\label{OceanpH}}
\end{figure}

An especially dramatic concern is that climate change may be acidifying oceans which is being blamed for threatening the future of coral reefs \cite{Anthony}.  Acidity is measured using the pH scale.  Distilled water is neutral and has a pH of 7.  Substances such as baking soda, ammonia and bleach have a pH $>$ 7 and are called bases while coffee, orange juice and stomach acid have a pH $<$ 7 and are called acids.  Sea water is slightly basic due to various impurities, most notably salt while rain is acidic.

\begin{figure}\centering
\includegraphics[height=100mm,width=.7\columnwidth]{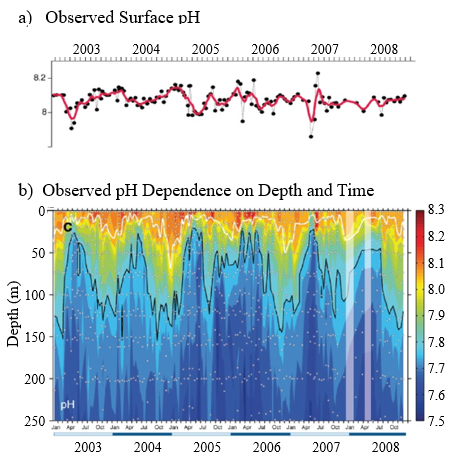}
\caption{Observations of Ocean Acidity in Santa Monica Bay during 2003 to 2008 a) pH averaged over the top 20 meters b) pH as function of time and depth \cite{Leinweber2013}.  The pH increases with depth because little sunlight penetrates the ocean to depths greater than 50 meters.  Hence, ocean plant life which extracts CO$_2$, is unable to flourish.  The CO$_2$ concentration increases with ocean depth and the pH decreases.  The seasonal and depth changes in ocean pH exceed the reduction of 0.27 that would occur if atmospheric CO$_2$ doubled from 400 to 800 ppm.
\label{MontereyPH}}
\end{figure}

About half of the carbon dioxide emitted into the air by burning fossil fuels is believed to be absorbed into the oceans \cite{Sabine}.  A small fraction of these carbon dioxide molecules react with sea water to generate carbonic acid \cite{Soli}.  The ocean acidity dependence on the atmospheric concentration of CO$_2$ has been calculated and is shown in Fig. \ref{OceanpH} \cite{Cohen}. It is the presence of carbon dioxide that prevents seawater from being very basic and makes the oceans habitable.  Doubling CO$_2$ from 400 to 800 ppm, reduces the seawater pH by 0.27.  This is less than changes as high as 1 pH per day that occur in some coastal regions due to the influx of inland fresh water caused by tides.  No adverse impact on the ecosystem has been detected  \cite{Hofmann}.  Fig. \ref{MontereyPH} shows the observed pH dependence on depth and time in Santa Monica Bay during 2003 to 2008.  A sunbathing seal who decides to dive 50 meters to catch a fish experiences a pH change comparable to that resulting from doubling atmospheric CO$_2$.  Similarly, the pH decrease due to the increase of CO$_2$ since the start of the Industrial Revolution is only 0.1.  

Ontario's contribution to ocean acidification can be estimated by multiplying the province's fraction of global CO$_2$ emissions, 0.35\%, by the reduction in pH, 0.27, that would result if atmospheric CO$_2$ were to double.  A CO$_2$ doubling time of 166 years is found by dividing the 2021 CO$_2$ concentration of 415 ppm by the observed rate of increase of 2.5 ppm/year.  This gives Ontario's contribution to ocean acidificiation of $6 \times 10^{-6}$ pH/year. 

\subsection{Precipitation}

The maximum amount of water vapour in the atmosphere increases 6\% per degree Celsius.  It is therefore reasonable that precipitation should increase in a warmer world \cite{Wentz}.  Indeed, it is hot humid summer days that produce torrential downpours.  The 2007 IPCC report stated that precipitation has increased in some regions by as much as 1\% in each decade of the 20th century \cite {IPCC2007, Dai}.  Several studies have examined precipitation for the last part of the 20th century.  The results range from a globally averaged precipitation trend that has changed by:  -1 mm/year \cite{Xie}; +3.5 mm/year \cite{Kistler}; and +0.1 mm/year \cite{Adler}.  These differences are not entirely surprising given that precipitation varies considerably over time scales of decades \cite{WvWPrecip}.  Data is also very sparse for large regions of the Earth including the Sahara, Amazon, Oceans etc. over most of the 20th century.  Hence, the resulting trends frequently are not statistically significant.  

\begin{figure} \centering
\includegraphics[height=60mm,width=.6\columnwidth]{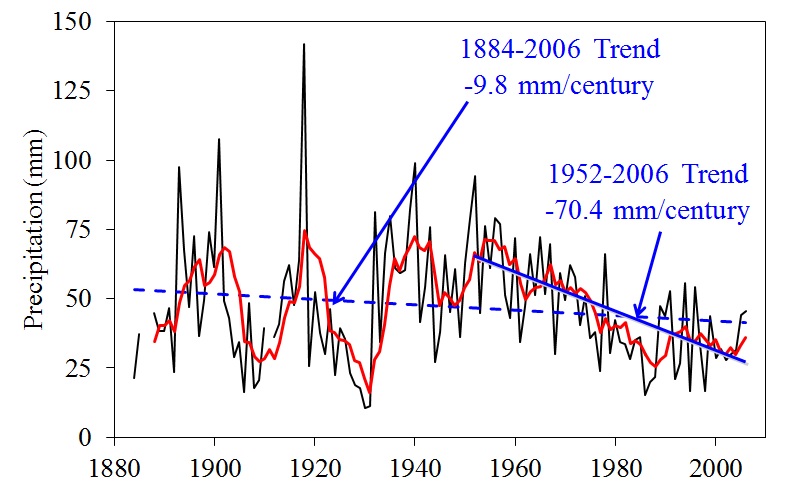}
\caption{Winter Precipitation in Medicine Hat, Alberta from 1880 to 2006. The data were downloaded from the Environment Canada Climate Archive \cite{EnvCan}.  The red line is the 5 year running average.  Large fluctuations of precipitation on timescales of decades are common in relatively dry regions of the Earth.  Precipitation trends can vary greatly for different time periods, even when extended times of half a century are considered.
\label{MedicineHatPrecip}}
\end{figure}

The need to be cautious about concluding precipitation has changed significantly is illustrated in Fig. \ref{MedicineHatPrecip}.  This shows winter precipitation at Medicine Hat, Alberta.  There is a sharply decreasing trend of -70.4 mm/century for the period 1952-2006, but the trend is much less if one considers data extending back to 1884.  The 1930s were particularly dry.  The central plains of North America were affected by a terrible drought and the region was called the dustbowl.  The large fluctuations of precipitation on timescales of years to decades is common in relatively dry areas.  

\begin{figure} \centering
\includegraphics[height=100mm,width=.5\columnwidth]{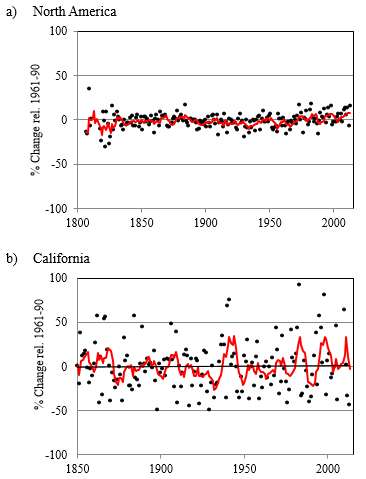}
\caption{Percentage Precipitation Change relative to 1961-1990 for a) North America and b) California \cite{WvWPrecip}.  The red line is the running 5 year average of the data.  There is no evidence of any long term significant change in precipitation for either North America or California.   The interannual variability of rainfall is much greater for a relatively dry region such as California than for stations located throughout North America.  
\label{NAPrecip}}
\end{figure}

Fig. \ref{NAPrecip} shows the percentage change in precipitation relative to 1961-1990 over the past 200 years for North America and California.  The interannual variability of precipitation when averaged over a large area such as North America is much smaller than when one considers a drier region like California.  California often experiences headline making droughts that the Governor has attributed to climate change \cite{Brown}.  Fig. \ref{NAPrecip} gives some pause to that claim.  

\begin{figure} \centering
\includegraphics[height=70mm,width=.6\columnwidth]{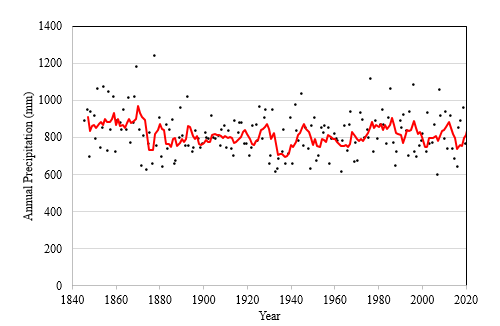}
\caption{Annual Precipitation in Toronto from 1843 to 2020 downloaded from the Environment Canada Archive \cite{EnvCan}.  The red line is the running 5 year average of the data.  The data fluctuate about the average annual precipitation value of 814 mm per year and show no significant trend. 
\label{TorontoAnnPrecip}}
\end{figure}

Fig. \ref{TorontoAnnPrecip} shows the annual precipitation at Toronto, Ontario from 1843 to 2020.  There is no statistically significant trend.  Ontario is blessed with a rather moderate climate not subject to prolonged multiyear droughts or floods.

The data shown in Figures \ref {NAPrecip} and \ref{TorontoAnnPrecip}  do not address whether once in a century downpours are becoming more common.  It is difficult to study the frequency of such events because several centuries of data are needed to make any meaningful conclusions.  Data for such extended times do not exist for Ontario.  That is not to say that heavy rainfall events do not produce greater flooding than in the past.  Such flooding, with extensive associated damage, invariably occurs in urban settings where the land surface has been greatly modified to accommodate concrete roadways, parking lots, buildings, etc. that make the surface impervious to water \cite{Bronstert}.  Whereas before it was developed, the land was able to soak up much rain like a sponge, the water now runs off into creeks turning them into raging torrents.  That may be a serious problem, but not one caused by climate change.  

\begin{figure} \centering
\includegraphics[height=100mm,width=1\columnwidth]{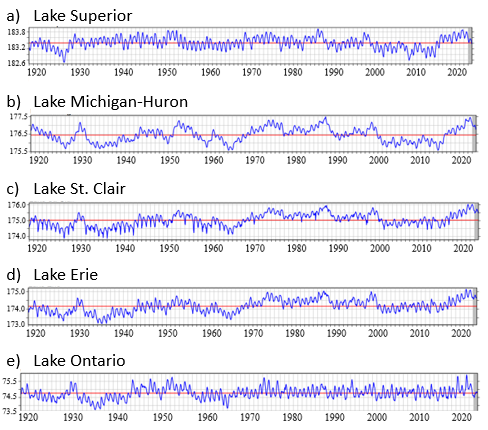}
\caption{Great Lake Levels relative to sea level as measured in meters from 1920 to 2020 by the U.S. Army Corps of Engineers, Great Lakes Hydraulics and Hydrology \cite{GreatLakeLevels}.  There is no evidence of any long term trend in Great Lake Levels which would change the incidence of flooding.
\label{GreatLakeLevels}}
\end{figure}

The Great Lakes watershed is the most important for Ontario as it contains the province's largest metropolitan areas.  Fig. \ref{GreatLakeLevels} shows the water levels of the Great Lakes relative to sea level measured by the U.S. Army Corps of Engineers from 1920 to 2020 \cite{GreatLakeLevels}.  The water levels depend not only on the amount of precipitation but also are affected by various dams comprising the St. Lawrence Seaway.  Even so, there is no trend for any Great Lake.  This supports the conclusion that in Ontario neither precipitation nor flooding due to high Great Lake water levels have changed significantly over the past century. 

\subsection{Extreme Events}

\subsubsection{Hurricanes}

A claim of the IPCC is that climate change is warming the oceans which has increased the severity and frequency of powerful storms.  This would make sense for hurricanes that develop in tropical regions and get their energy from warm surface water.  However, this prediction may be simplistic.  An important driver of the Earth's climate is the temperature difference betweeen the equatorial and polar regions.  Weather systems transfer energy away from the tropics.  Global warming will be largest in the Arctic.  Hence, the temperature difference between the Arctic and the equator will be reduced, with presumably a subsequent weaking or shift of wind patterns such as the Jet Stream \cite{Zanchettin}.  Hurricanes, with the exception of Hurricane Hazel in 1954, seldom strike Ontario.  

\begin{table}
\begin{center}
\begin{tabular}{|c|c|}
\hline
Hurricane &Sustained Windspeed\\ [0.5ex]
Category &km/hour\\
\hline\hline
1  &119-153\\
\hline
2  &154-177\\
\hline
3 &178-208\\
\hline
4 &209-251\\
\hline
5 &$\ge 252$\\
\hline
\end{tabular}
\end{center}
\caption{Saffir Simpson Hurricane Wind Scale \cite{Safir}.
\label{acd2}}
\end{table}

\begin{figure} \centering
	\includegraphics[height=80mm,width=.8\columnwidth]{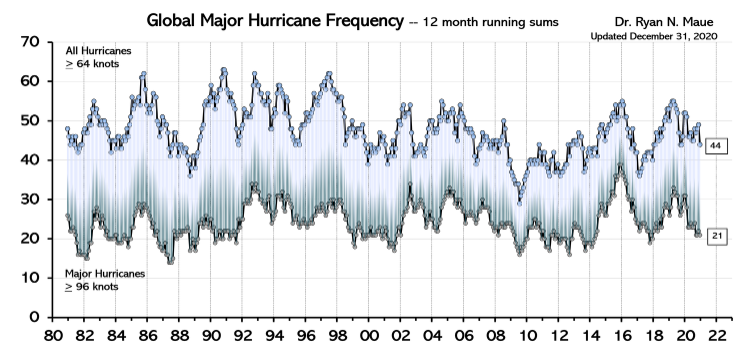}
	\caption{Annual Number of Global Hurricanes Observed from 1981 to 2021.  No trend is evident in either the total number of hurricanes or the number of major hurricanes \cite{HurricanesMaue},\cite{HurricanesPielke}. 
		\label{HurricanesMaue}}
\end{figure}

The amount of hurricane damage very strongly depends on the maximum sustained wind speed.  Hurricanes are categorized using the Saffir Simpson wind scale given in Table \ref{acd2}.  Damage to well constructed homes can be major even for a Category 1 hurricane.  Large tree branches may break and broken power lines may result in electrical outages for several days.  Category 5 hurricanes may make an area uninhabitable for months.  Fig. \ref{HurricanesMaue} shows the number of global hurricanes observed from 1981 to 2021.  No obvious trend is evident in either the total number of hurricanes or the number of major hurricanes defined as having wind speed in excess of 96 knots (153 km/hr).

\subsubsection{Tornadoes}
Another extreme weather event is tornadoes.  Most of the world's tornadoes occur in the United States.  They originate when a warm air front from the Gulf of Mexico collides with cooler air from Northern Canada.  The damage caused by a tornado depends on the wind strength and is categorized by the Enhanced Fujita scale shown in Table \ref{Tornado}.

\begin{figure} \centering
\includegraphics[height=100mm,width=0.5 \columnwidth]{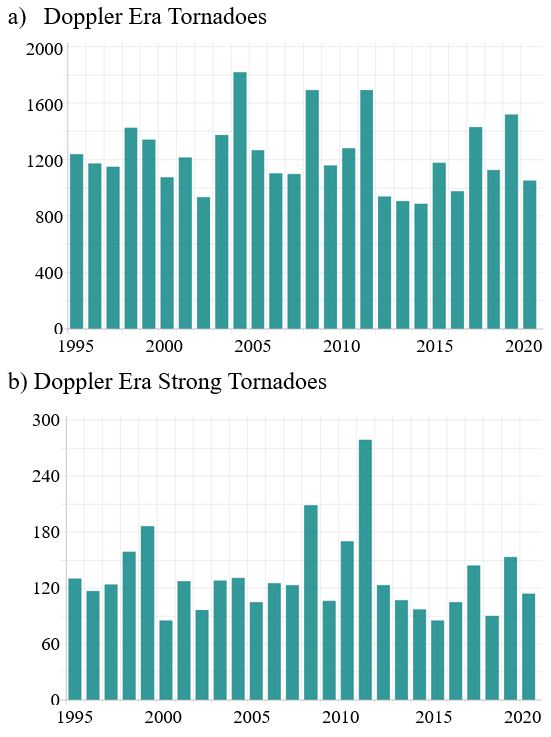}
\caption{Doppler Era U.S. Tornadoes from 1995 to 2020 for a) All Tornadoes and b) Strong (with rating of EF2 or greater) Tornadoes \cite{USTornadoes}.  No trend is evident in either the total number of tornadoes or the number of strong tornadoes.
\label{USTornadoes}}
\end{figure}

The number of tornadoes recorded has increased over the past 50 years as improved technology, especially Doppler radar, has been introduced \cite{USTornadoes}.  Fig. \ref{USTornadoes} shows the number of tornadoes observed in the U.S. from 1995 to 2020.  There is no significant trend in either the total number of tornadoes or the number of strong tornadoes, defined as having a rating of EF2 or greater.   

\begin{table}
	\begin{center}
		\begin{tabular}{|c|c|}
			\hline
			EF Rating &3 Second Wind Gust\\ [0.5ex]
			&km/hour\\
			\hline\hline
			0 &90-130\\
			\hline
			1 &135-175\\
			\hline
			2 &180-220\\
			\hline
			3 &225-265\\
			\hline
			4 &270-310\\
			\hline
			5 &$>$ 315\\
			\hline
		\end{tabular}
	\end{center}
	\caption{Enhanced Fujita Tornado Scale \cite{Fujita}.
		\label{Tornado}}
\end{table}

\begin{figure} \centering
\includegraphics[height=100mm,width=.9\columnwidth]{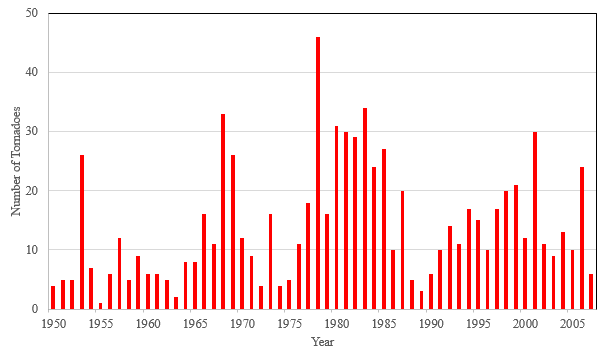}
\caption{Annual Number of Tornadoes Observed in Ontario from 1950 to 2007 \cite{OntTornadoes}. The higher number of tornadoes observed after 1978 is very likely due to better observing techniques including the use of improved technology such as Doppler radar.  The data appear to show a decreasing trend after 1978.
\label{OntTornadoes}}
\end{figure}

Fig. \ref{OntTornadoes} shows the number of tornadoes recorded in Ontario from 1950 to 2007 \cite{OntTornadoes}.  There were fewer tornadoes in the period 1950 to 1975 than in later years.  This is readily explained by the introduction of better observing technology.  There seems to be a decreasing trend after 1978.    

\subsubsection{Forest Fires}

\begin{figure}\centering	
\includegraphics[height=120mm,width=.6\columnwidth]{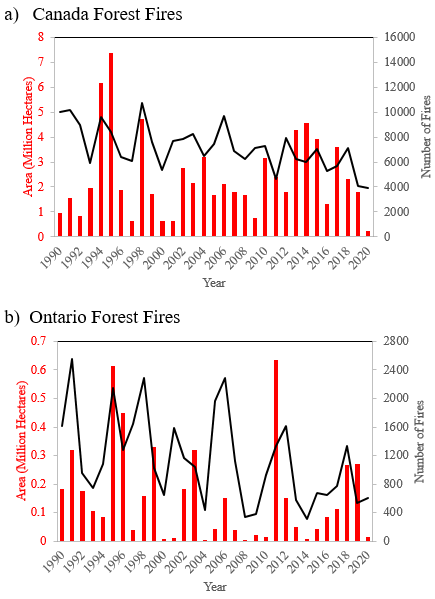}
\caption{Area Burned and Number of Forest Fires from 1990 to 2020 in a) Canada and b) Ontario according to Canada's National Forestry Database \cite{CanOntForestFires}.  The number of forest fires has decreased for both Canada and Ontario during the past 30 years while the area burned has not changed substantially. 
\label{CanOntForestFires}}
\end{figure}

A significant negative effect of decreasing precipitation and/or higher temperatures would be an increased risk of forest fires.  It should be pointed out that forest fires caused by lightning are a natural part of the ecosystem.  The concern is that climate change is increasing drought severity making fires more numerous and larger.  Fig. \ref{CanOntForestFires} shows a small decrease in the number of wildfires in Canada and Ontario from 1990 to 2020.  There is no clear trend in the number of hectares burned.  This is somewhat surprising given that mankind has strongly interfered with the forest ecosystem.  Policies to extinguish fires have allowed a build up of flammable material in North America's forests during the 20th century.  Hence, one would expect fires to be much larger.  In addition, man has all too frequently been careless, inadvertently starting fires.

\newpage
\section{Summary}
	
Table \ref{Summary} summarizes the evidence.  In 2019, Ontario generated only 0.35\% of global CO$_2$ emissions.  This amount was 20\% lower than in 2005.  This record of greenhouse gas reduction is nearly unparalleled in the world! 

\begin{table}
	\begin{center}
		\begin{tabular}{|c|l|}
\hline
&\\
			Subject &\ \ \ \ \ \ \ \ \ \ \ \ \ \ \ \ \ \ \ \ \ \ \ \ \ \ \ \ \ \ \ \ \ \ \ \ \  Status\\ [0.5ex]
&\\
\hline\hline
&\\
Greenhouse &-Ontario's CO$_2$ equivalent emissions in 2019 were only 0.35\% of world total. \\
 Gas                        &-Ontario emissions in 2019 were 20\% lower than in 2005.\\
                         &-Almost no other country has matched Ontario's greenhouse gas reduction! \\
&\\
\hline
&\\
Temperature &-Global average temperature has increased by about 1$^o$ C since 1880.\\
            &-Climate models do not account for decadal temperature fluctuations.\\
            & i.e. warming before 1940, slight cooling 1940 - 1980, hiatus 2000 - 2016.\\
            &-GCMs overestimate warming by about a factor of 2.\\ 
            &-Ontario's global warming contribution is $9.2 \times 10^{-5}$ $^o$C/year using $S=3$ $^o$C.\\
&\\
\hline
&\\
Polar &-Reduction in Arctic ice 1979-2022 consistent with temperature increase.\\
Ice Caps		  &-Minimum Extent of Arctic ice in September stabilized after 2007.\\
		  &-Summer Arctic ice likely to be around for centuries.\\
          &-Slight increase in average Antarctic ice cap during 1979-2022. \\
&\\
\hline
&\\
Oceans &-Sea level increased by 100 meters at end of last ice age.\\
       &-Ontario responsible for anthropogenic sea level increase of only 0.005 mm/year.\\
       &-Ontario Hudson Bay sea level declining due to isostatic rebound.\\
       &-Ocean acidification less than pH change due to tides, depth \& season.\\
       &-Ontario's contribution to ocean acidification less than $6 \times 10^{-6}$ pH/year.\\
&\\		
\hline
&\\
Precipitation&-No clear evidence of change in long term precipitation.\\
	         &-Toronto precipitation very stable since 1843.\\
	         &-No evidence of more floods as shown by stable Great Lake levels from 1920 - 2020.\\
&\\
\hline
&\\
Extreme &-No change in number or severity of hurricanes during 1981 - 2021.\\
Events              &-No change in number or severity of U.S. tornadoes during 1995 - 2020.\\
               &-Evidence of decline in number of Ontario tornadoes from 1978 - 2005.\\
               &-Decrease in number of forest fires in Canada \& Ontario during 1990 - 2020.\\
&\\
\hline
		\end{tabular}
	\end{center}
	\caption{Summary of Global Warming Effects Relevant to Ontario}.
		\label{Summary}
\end{table}

The Earth's average surface temperature has increased by about 1 $^o$C since 1880.  Global climate models are not able to account for substantial decadal variations.  The most significant of these is the abrupt warming that occurred in the 1990s followed by the so called hiatus from 2000 to about 2016.  A comparison with observations shows a consistent overestimation of the warming by nearly all GCMs. 

The 0.5 $^o$C warming from 1900 to 1940 at relatively low CO$_2$ levels show natural temperature fluctuations are a substantial part of the overall 1$^o$C warming from 1880 to 2020.  It is reasonable to conclude that about half of the temperature rise since 1880 is due to increasing greenhouse gases.  Substituting 0.5 $^o$C into equation (1) and using the corresponding CO$_2$ concentrations in 1880 and 2020 gives a climate sensitivity $S=1$ $^o$C.  This is less than the 2021 IPCC climate sensitivity range of 2.5 to 4 $^o$C.  If one attributes all of the observed 1 $^o$C warming from 1880 to 2020 to increasing greenhouse gases, the climate sensitivity is only 2 $^o$C.  

Ontario's contribution to global warming can be estimated using its fraction of world CO$_2$ production.  Using the 2021 IPCC recommended climate sensitivity, $S=3$ $^o$C, gives a contribution of $9.2 \times 10^{-5}$ $^o$C or 92 microdegrees C per year to global warming.  This is dwarfed by that of other countries, especially China, U.S. and India.  

There has been a significant reduction in the Arctic ice cap since satellite measurements began in 1979.  However, the extent of minimum sea ice in September appears to have stabilized after 2007.  Extrapolating the 1979-2022 data to determine when there will be an ice free Arctic Ocean gives a date of 2077.  Using the 2007-2022 data changes the ice free year to 4729.  Records of Greenland suggest that the warming of the last few years is less than that experienced in the first part of the 20th century.  The dramatic retreat of glaciers such as in Glacier Bay, Alaska during the 1800s and early 1900s occurred when greenhouse gas emissions were relatively low and is likely due to warming fo the Earth following the end of the Little Ice Age.  For Antarctica, there has been a slight increase in average sea ice extent since 1979 although a slight reduction in recent years.

Sea levels have continued increasing throughout the 20th century at a rate of 2 to 3.5 mm/year.  This is likely a continuation of a natural process that began at the end of the last ice age.  Ontario's contribution to anthropogenic global sea level rise is only 0.005 mm/year.  For Ontario's sea coast bordering Hudson Bay, sea level will likely continue a significant decline due to isostatic rebound of the land in response to the end of the last ice age.  The acidity change due to absorption of CO$_2$ by the oceans is small compared to tidal and season effects as well as the pH change experienced when one dives below the surface.  Ontario's contribution to ocean acidification is a minuscule $6 \times 10^{-6}$ pH/year.  
  
North American precipitation records show no change over the past two centuries.  Precipitation recorded at Toronto from 1843 to 2020 is remarkably stable.  Over a century of measurements show no significant change of flooding caused by high Great Lake water levels.  

The frequency of extreme events was examined.  No change in either the number of total hurricanes or severe hurricanes during 1981 to 2021 was found.  Similarly, tornado observations using modern Doppler radar show no change for either the number of total tornadoes or strong tornadoes from 1995 to 2020 in the U.S.  The number of Ontario tornadoes shows no overall trend from 1950 to 2007 but there appears to be a decrease from 1978 to 2007.  The annual number of forest fires decreased in Canada and Ontario from 1990 to 2020 and the area burned shows no clear trend.

In conclusion, the author wishes to emphasize that application of well tested physics shows an increase of greenhouse gases will cause some global warming.  However, contrary to the United Nations claim, there will not be catastrophic climate change within 10 years \cite{Garces}, nor should one overlook the beneficial aspects of increased carbon dioxide for promoting plant growth.  The impact of Ontario's anthropogenic emissions of greenhouse gases on global climate is minuscule. 

\section*{Acknowledgements}
We are grateful for the financial support of the Canadian Natural Science and Engineering Research Council.

\end{document}